\documentclass[aps,prb,twocolumn,showpacs,amsmath,amssymb,superscriptaddress,floatfix]{revtex4-1}
\usepackage{graphicx}
\usepackage{dcolumn}
\usepackage{bm}
\usepackage{amsfonts}
\usepackage{amsmath}
\usepackage{ulem}
\usepackage{color}
\usepackage{hyperref}
\begin{document}

\title{Effects of disorder on electron tunneling through helical edge states} 

\author{Pietro Sternativo}
\affiliation{Dipartimento di Scienza Applicata e Tecnologia del Politecnico di Torino, I-10129 Torino, Italy}
\author{Fabrizio Dolcini}
\email{fabrizio.dolcini@polito.it}

\affiliation{Dipartimento di Scienza Applicata e Tecnologia del Politecnico di Torino, I-10129 Torino, Italy}

\affiliation{CNR-SPIN, Monte S.Angelo - via Cinthia, I-80126 Napoli, Italy}

\begin{abstract}
A tunnel junction between helical edge states, realized via a constriction in a Quantum Spin Hall system, can be exploited to steer both charge and spin current into various terminals. We investigate the effects of disorder on the transmission coefficient $T_p$ of the junction, by modelling disorder with a randomly varying (complex) tunneling amplitude $\Gamma_p=|\Gamma_p| \exp[i \phi_p]$.  We show that, while for a clean junction $T_p$ is only determined by the absolute value $|\Gamma_p|$ and is independent of the  phase $\phi_p$, the situation can be quite different in the presence of disorder: phase fluctuations may dramatically affect the energy dependence of $T_p$ of any single sample. Furthermore, analysing three different models for phase disorder (including correlated ones), we show that not only the amount but also the way the phase $\phi_p$ fluctuates determines the localization length~$\xi_{loc}$ and the sample-averaged transmission. Finally, we discuss the physical conditions in which these three models suitably apply to realistic cases. 
\end{abstract}

\pacs{73.23.-b, 73.43.Jn, 71.23.-k}

\maketitle

%%%%%%%%%%%%%%%%%%%%%%%%%%%%%%%%%%%%%%%%%%%%%%%%%%%%%%%%%%%%%%%%%%%%%%%%%%%%%%%%%%%%
%%%%%%%%%%%%%%%%%%%%%%%%%%%%%%%%%%%%%%%%%%%%%%%%%%%%%%%%%%%%%%%%%%%%%%%%%%%%%%%%%%%%
%%%%%%%%%%%%%%%%%%%%%%%%%%%%%%%%%%%%%%%%%%%%%%%%%%%%%%%%%%%%%%%%%%%%%%%%%%%%%%%%%%%%
\section{Introduction}
Theoretical predictions~\cite{TM-1-theo} and experimental evidence~\cite{TM-1-exp} have shown that the one-dimensional conducting  channels emerging at the edges of a Quantum Spin Hall Effect (QSHE) system are {\it helical}, so that along one boundary  spin-$\uparrow$ electrons propagate (say) rightwards, and spin-$\downarrow$ electrons  leftwards.\cite{hasan-review,zhang-review}

A spectacular effect  connected to such helical nature is the topological protection from scattering off non-magnetic impurities, which ideally makes helical edge states perfectly conducting 1D channels. This property has inspired various investigations about the effects of disorder on helical states. On the one hand, various studies have tested such robustness to disorder when inelastic scattering is included~\cite{TM-1-theo,moore_2006,zhang-PRL,oreg_2012,glazman_2013}, possibly in interplay with Rashba impurities\cite{johannesson_2010,crepin_2012,budich_2012,schmidt_2012,geissler_2014}, or when time-reversal symmetry is broken either by magnetic impurities~\cite{cheianov-glazman,maciejko_2009,tanaka_2011,sassetti_2014} or by applied magnetic fields.~\cite{buettiker_2012,wimmer_2014,sun_2014}  On the other hand, it also been realised that disorder itself can cause an ordinary insulator to undergo a phase transition  to a topological insulator (Topological Anderson Insulator).~\cite{shen_2009,jiang_2009,beenakker_2009,chen_2012}  In turn, such transition may be strongly modified when disorder exhibits  spatial correlations.~\cite{rotter}  

An important consequence of the helical property is that QSHE edge states represent a promising platform for applications to spintronics.\cite{spintronics-TI} In particular, it has been predicted that charge and spin currents can be steered in multi-terminal devices exploiting a tunnel coupling between the four edge states~\cite{teo,chamon_2009,strom_2009,akhmerov,trauz-recher,virtanen-recher,dolcini2011,richter,citro-romeo,citro-sassetti,dolcini2012,noise,dolcetto-sassetti,sassetti-cavaliere-2013,liliana,ferraro,sternativo,schmidt_2013,buettiker_2013}. Typically, such tunnel coupling is modelled as a clean quantum point contact.  
 However, various sources of disorder may arise when creating a constriction in a QSHE quantum well. 
Indeed, due to the Klein tunneling characterizing the Dirac spectrum of the helical states, simple gating would not be effective to create a constriction, which instead has to be realized through lithographic techniques combined with an etching process. In doing that, oxides tend to form at the border of the etched region, leading to a randomly varying electric potential. In addition, the geometrical profile of the constriction may exhibit roughness, resulting in a randomly varying width of the constriction. Furthermore,  a top gate is needed to drive the system into the QSHE regime, and the dielectric layer separating the metallic gate from the quantum well is typically an amorphous insulator that naturally leads to local potential fluctuations.   Disorder is thus an intrinsic feature in helical tunnel junctions that has to be accounted for.

%%%%%%%%%%%%%%%%%%%%%%%%%%%%%%%
\begin{figure} 
\centering
\includegraphics[width=8cm,clip]{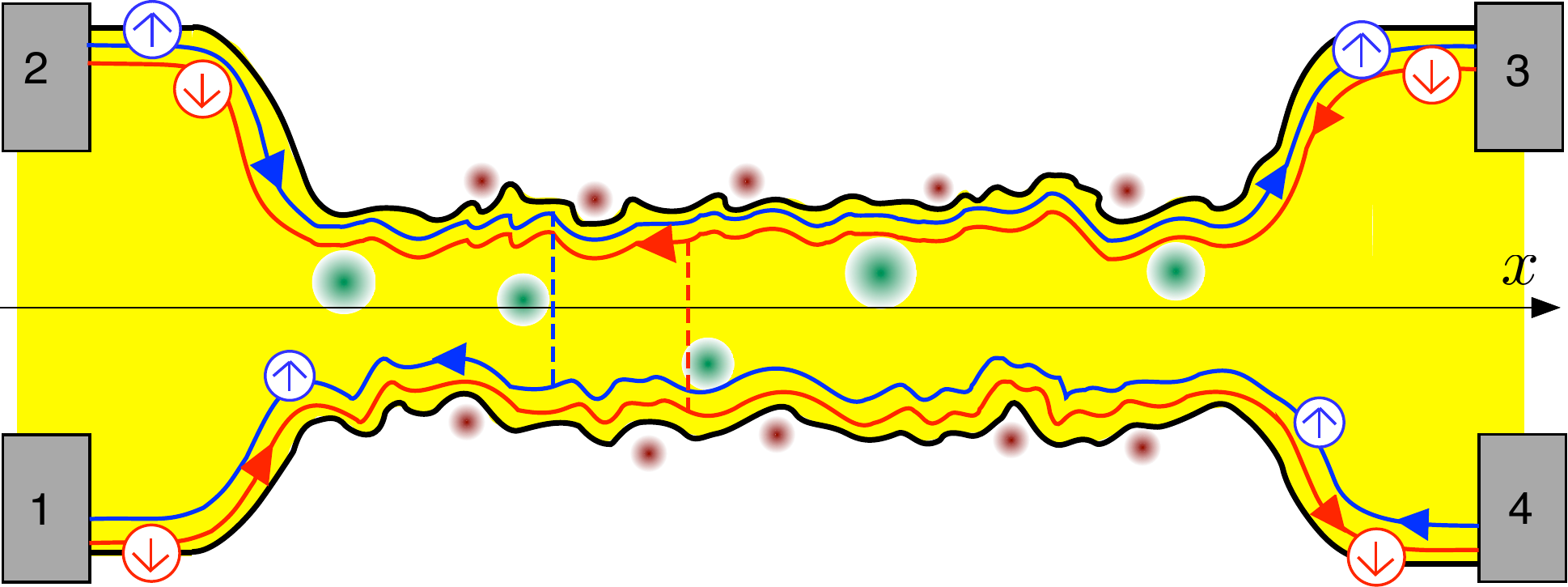}
\caption{(Color online) A constriction etched in a four-terminal Quantum Spin Hall effect setup induces an electron tunnel coupling between the helical edge states. Vertical   arrows inside the circles indicate electron spin orientation. A fluctuating width along the longitudinal direction $x$ causes geometrical disorder, while electrical disorder can originate from the presence of oxides (small red dots) produced by the etching process and/or from the potential fluctuations (bigger green dots) due to the amorphous dielectric separating the conducting channel from the top gate.}
\label{Fig-01}
\end{figure}
%%%%%%%%%%%%%%%%%%%%%%%%%%%%%%%%%%%%%%%%%%%
So far, analytical predictions about disorder effects are mostly available for short tunnel junctions. In that limit, the spin texture of the helical states  has been shown to strongly affect the localization length,~\cite{schmidt_2013}  whereas  the presence of a magnetic field, whose breaking of time-reversal symmetry induces backscattering along the edge states, leads to a peak   in the noise correlations.~\cite{buettiker_2013}
As far as finite length tunnel junctions are concerned, results about disorder are mostly limited to numerical approaches that analyzed the transmission coefficient of individual samples.\cite{richter}    
A thorough investigation covering both single-sample and intrinsic disorder properties of helical tunnel junctions is still lacking. Furthermore, already in the time-reversal symmetric case, it would be desirable for a comparison with experiments to analyze how and to what extent the various disorder sources affect the transmission coefficient.

In this paper we address these problems, analyzing the effects of disorder on the QSHE setup depicted in Fig.\ref{Fig-01}, 
using an effective 1D model, presented in Sec.\ref{sec-II}, that assumes a randomly varying (complex) electron tunneling amplitude between the helical edge states. We first analyze  in Sec.\ref{sec-III} the properties of individual samples, and show that the disorder on the {\it absolute value} and on the {\it phase} of the tunneling amplitude may have quite different effects on the transmission coefficient of a disordered sample. In particular, while in a clean tunnel junction the phase is unimportant, for a disordered junction it plays a major role in realistically relevant regimes.  
We then address  in Sec.\ref{sec-IV}  the localization length, the sample-independent intrinsic disorder property that determines the lengthscale over which the transmission decays as a function of the sample length. We show that not only the typical amount of phase fluctuations, but also the {\it way} the phase fluctuates strongly affects the localization length $\xi_{loc}$. In particular, we analyze three different models for phase fluctuations, including the case of correlated disorder, and show three different behaviours for the predicted energy dependence of $\xi_{loc}$. Also, we discuss how these differences impact the sample-averaged transmission. Finally, in   Sec.\ref{sec-V} we discuss how the proposed models apply in realistic implementations of helical tunnel junctions, and compare our results to the case of a conventional disordered quantum wire.

%%%%%%%%%%%%%%%%%%%%%%%%%%%%%%%%%%%%%%%%%%%%%
%%%%%%%%%%%%%%%%%%%%%%%%%%%%%%%%%%%%%%%%%%%%% 
%%%%%%%%%%      M O D E L    %%%%%%%%%%%%%%%%
%%%%%%%%%%%%%%%%%%%%%%%%%%%%%%%%%%%%%%%%%%%%% 
%%%%%%%%%%%%%%%%%%%%%%%%%%%%%%%%%%%%%%%%%%%%% 
 
\section{Model}
\label{sec-II}

For a long and clean constriction it has been shown~\cite{zhou,richter} that, starting from the 2D Bernevig-Hughes-Zhang (BHZ) model defined on a stripe of width $W$, one can obtain an effective 1D model defined on the basis of the edge modes, which turn out to be coupled via a tunneling amplitude related to their wavefunction overlap. Such an overlap roughly decays exponentially with the transversal width $W$ of the junction.  A straightforward way to account for the geometrical disorder due to the profile roughness is to adopt such a basis, and to assume  a width~$W(x)$ that  fluctuates along the longitudinal direction $x$,  resulting in a randomly varying tunneling amplitude. 
Notice that, as is well known, there are in fact two types of tunneling amplitudes, $\Gamma_p(x)$ and $\Gamma_f(x)$, related to spin-preserving(p) and spin-flipping(f) processes, respectively, both ensuring time-reversal symmetry.~\cite{zhang-PRL,teo,richter,trauz-recher,citro-romeo,dolcini2011,citro-sassetti,sternativo,SO-HgTe,dellanna} The model presented here can in principle include both, although we shall be mainly interested in the effects of the first coupling $\Gamma_p$, which is expected to be dominant. \\
Furthermore, the disorder arising from the oxides formed when etching the constriction and from the inhomogeneities of the amorphous dielectric can be regarded as a series of randomly distributed electric potential centers, each coupling to the edge state densities  in two different ways, $V_T(x)$ and $V_B(x)$, depending on the transversal distance from the top and bottom edges, and   randomly varying along the longitudinal direction~$x$. \\
 
We describe the helical edge states   by four  electron field operators, namely 
$\Psi_{R\uparrow}(x) ,
\Psi_{L\downarrow}(x)$ for the (say) upper edge and $\Psi_{R\downarrow}(x) ,
\Psi_{L\uparrow}(x)
$ for the lower edge, characterized by a linear spectrum Dirac Hamiltonian~\cite{TM-1-theo,TM-1-exp},  
\begin{equation}
\hat{\mathcal{H}}_0 = -i    \hbar v_{\rm F} \hspace{-0.35cm} \sum_{\alpha= R/L=\pm} \hspace{-0.2cm}\alpha \sum_{\sigma=\uparrow, \downarrow}\int   \! dx     :   \Psi^{\dagger}_{\alpha \sigma}(x)\, \partial_x \Psi^{}_{\alpha \sigma}(x)    : \quad,                
\label{H0}
\end{equation}
with $\alpha=R/L=\pm$ denoting the chirality for right- and left- movers, respectively,~\cite{footnote-RL-pm} $\sigma=\uparrow,\downarrow$ is the spin component, and $: \, \, \, :$ indicates the normal ordering.
%%%%%%%%%%%%%%%%%%%%%%%%%%%%%%%%%%%%%%%%%%%%%%%%%
The most general tunneling terms that preserve time-reversal symmetry are given by 
\begin{eqnarray}
\hat{\mathcal{H}}_{\rm tun}  &=& \! \!  \displaystyle \sum_{\sigma=\uparrow, \downarrow}\! \int   \! dx      \left( \Gamma_{p}^{}(x) \,  \Psi^{\dagger}_{L \sigma}(x)\, \Psi^{}_{R \sigma}(x)    +   \mbox{H.c.} \right) \, \;  
\label{Hsp} \\
& &   +  \hspace{-0.3cm}\sum_{\alpha=R/L=\pm} \hspace{-0.3cm} \alpha \int   \! dx    \,  \left(   \Gamma_{f}^{}(x) \, \,  \Psi^{\dagger}_{\alpha \downarrow}(x)\, \Psi^{}_{\alpha \uparrow}(x)   \, +     \mbox{H.c.} \right)   \quad,  \nonumber   
\end{eqnarray}
where $\Gamma_p(x)$ and $\Gamma_f(x)$ describe the randomly varying tunneling amplitudes related to spin-preserving and spin-flipping processes, respectively~\cite{zhang-PRL,teo,richter,trauz-recher,citro-romeo,dolcini2011,citro-sassetti,sternativo}.   
The coupling with a potential that randomly fluctuates in space reads
\begin{eqnarray}
\hat{\mathcal{U}}  &=&  \displaystyle  \! \int    \! dx   \left[  \, eV_{T}(x) \left(   \hat{\rho}_{R \uparrow}(x)\,     \,+  \hat{\rho}_{L \downarrow}(x)\,  \right) \right. \label{U}  \\
  & &  \left. \displaystyle  \hspace{0.5cm} +eV_{B}(x) \left(   \hat{\rho}_{R \downarrow}(x)    \,+ \,  \hat{\rho}_{L \uparrow}(x)   \right) \right] \quad,
\nonumber 
\end{eqnarray}
where $
\hat{\rho}_{\alpha \sigma}(x)=\, :\Psi^\dagger_{\alpha \sigma}(x) \Psi^{}_{\alpha \sigma}(x):
$
is the electron chiral density.  Equation (\ref{U}) can also be rewritten as 
\begin{eqnarray}
\hat{\mathcal{U}}  &=&  \displaystyle  \! \int    \! dx   \left[  \,  V_{p}(x) \, \hat{\rho}_{c}(x)  \, + \, \,  V_{f}(x) \,  \hat{\jmath}_{s}(x)/v_F \right] \label{U-2}  
\end{eqnarray}
where 
\begin{equation}
V_{p/f}(x) =  (V_{T}(x)\pm V_{B}(x)) \, / 2  \quad, \label{Vgcs-def}
\end{equation}
couple to the charge density $\hat{\rho}_c {\,=\,} e(\hat{\rho}_{R \uparrow}+\hat{\rho}_{L \uparrow}+\hat{\rho}_{R \downarrow} +\hat{\rho}_{L \downarrow})$  and to the spin current $\hat{\jmath}_s {\,=\,} e v_F (\hat{\rho}_{R \uparrow}+\hat{\rho}_{L \downarrow}-\hat{\rho}_{L \uparrow}-\hat{\rho}_{R \downarrow})$, respectively. 
The full Hamiltonian for the disordered junction thus reads 
\begin{equation}
\hat{\mathcal{H}} =\hat{\mathcal{H}}_{0}   \, + \,   
\hat{\mathcal{H}}_{\rm tun}   \, + \,
\hat{\mathcal{U}}     \quad.     
\label{HAM-FER}   
\end{equation}
%%%%%%%%%%%%%%%%%%%%%%%%%%%%%%
Far away from the constriction  the helical edge states propagate freely and are eventually injected or absorbed by the 4 metallic  electrodes (see Fig.\ref{Fig-01}), each kept at a chemical potential $\mu_i$ ($i=1,\ldots 4$). Denoting by $x_0$ and~$x_f$ the left and right extremal longitudinal coordinates of the constriction, the system is disorder-free ($\Gamma_{p}, \Gamma_{f}, V_{p}\, , V_{f} \, \equiv 0$) for $x < x_0$ and $x>x_f$, whereas within  the finite length $L=x_f-x_0$ of the constriction region  the tunneling amplitudes $\Gamma_{p}(x) \, , \Gamma_{f}(x)$ and the   potentials $V_{p}(x),V_{f}(x)$   fluctuate randomly along the longitudinal direction $x$.   \\

In Ref.[\onlinecite{sternativo}] it has been shown for model (\ref{HAM-FER}) that, even though the {p}-tunneling and {f}-tunneling terms in (\ref{Hsp}) do not commute, the four-terminal scattering matrix $\mathsf{S}$ as well as the related transconductance matrix $\mathsf{G}$ exhibit a factorisation into {p}- and {f}-processes. More specifically, each entry of $\mathsf{G}$ is factorized into a product of two terms, one depending on only the profile of $\Gamma_p(x)$ and $V_{p}(x)$, and the other depending on only $\Gamma_f(x)$ and $V_{f}(x)$ only. Thus, if a bias voltage $V$ is applied to, say, terminal 2 in Fig.\ref{Fig-01}, the resulting currents entering terminals 1, 3 and 4 are
\begin{eqnarray}
I^{(1)} &= & \, (1-T_p) \, \frac{{\rm e}^2}{h} \, V  \label{I1} \\
I^{(3)} &= &   T_p \, T_f\, \frac{{\rm e}^2}{h} \, V \, \label{I3} \\  
I^{(4)} &= &   T_p \, (1-T_f) \, \frac{{\rm e}^2}{h} \, V \, \quad.\label{I4}
\end{eqnarray}
Here
\begin{eqnarray}
T_{p}&=&    T_{p}[ \Gamma_{p}(x) ; V_{p}(x) ;E]   \label{Tp-dep}\\
T_{f}&=& T_{f}[\Gamma_{f}(x); V_{f}(x)]   \label{Tf-dep}  
\end{eqnarray}
are the transmission coefficients related to spin-preserving ({p}-sector) and spin-flipping  ({f}-sector) tunneling processes, respectively, and can be  operatively determined via Eqs.(\ref{I1})-(\ref{I3})-(\ref{I4}).\\
  
The separate dependence of $T_p$ and $T_f$, described by Eqs.(\ref{Tp-dep}) and (\ref{Tf-dep}), holds for any static profile of $\Gamma_{p/f}$ and $V_{p/f}$ that fluctuates randomly in space (elastic coupling). In the opposite limit, where an additional degree of freedom that is spatially localised and that dynamically couples to the edge states, inelastic backscattering can occur.~\cite{glazman_2013}

%%%%%%%%%%%%%%%%%%%%%%%%%%%%%%%%%%%%%%%%%%%%%%%%%%%%%%%%%%%%
\subsection{Eliminating the potentials $V_p(x)$ and $V_f(x)$ via a gauge transformation}
\label{sec-II-a}
{\it A priori}, in both sectors $\nu=p,f$, the model is characterized by three real parameters, namely a (complex) tunneling amplitude $\Gamma_\nu(x)=|\Gamma_\nu(x)| \exp[i \phi_\nu(x)]$ and a (real) potential  $V_{\nu}(x)$. In fact, only {\it two} parameters are sufficient. Indeed we show here below  that the potential  $V_{\nu}(x)$   can be reabsorbed into the phase $\phi_\nu$ of the  tunneling amplitude, via a gauge transformation $\phi_\nu(x) \rightarrow \phi^\prime_\nu(x)$, where the renormalised phases read 
\begin{equation}
\phi^\prime_{\nu}(x) {\,=\,} \phi_{\nu}(x)-2 \int_{x_0}^x \frac{eV_{\nu}(x^\prime)}{\hbar v_F} \, dx^\prime \hspace{0.7cm} \nu=p,f \quad.\label{gauge-phase}
\end{equation}
Thus, while the potential in Eqs.(\ref{U}) and (\ref{U-2}) alone does not couple the two edges directly, it does affect their tunneling term (\ref{Hsp}). 
Importantly, the phase  $\phi^\prime_{\nu}(x)$ depends {\it non-locally} on the related potential $V_\nu$.  In particular,  a constant potential $V_\nu(x) \equiv V_\nu$ yields a phase $\phi^\prime_{\nu}(x) \sim -2  eV_{\nu} x/\hbar v_F$ that varies {\it linearly} in space,   whereas a $\delta$-like potential $V_\nu(x) =V^0_\nu \delta(x-x_j)$ centered at $x_j$  yields a {\it phase-jump} at $x_j$, $\phi^\prime_{\nu}(x) \sim-2  eV^0_{\nu} \theta(x-x_j)/\hbar v_F$, where $\theta(x)$ is the Heaviside step function.
This implies that different profiles for the phase $\phi_\nu$ have to be adopted to account for various types of $V_\nu(x)$ disorder.~\cite{nota-gauge-inv}  \\

A suitable way to prove that the potentials $V_{\nu}(x)$ can be gauged away is to exploit the equation of motion for the electron fields obtained in Ref.[\onlinecite{sternativo}].  
For this purpose, we introduce a four component electron field operator
$
\Psi  =\left( 
\Psi_{R\uparrow}  ,
\Psi_{L\uparrow} ,
\Psi_{R\downarrow}  ,
\Psi_{L\downarrow}  
\right)^T
$, 
and rewrite the Hamiltonian (\ref{HAM-FER}) in a compact form as
\begin{eqnarray}
\mathcal{H} &=& \hbar v_F  \int dx \, \Psi^\dagger \left\{ -i \partial_x (\sigma_0 \otimes \tau_z) + \right. \label{HAM-FER-MAT}\\
& & \left. +\sigma_0 \otimes \tau_z [ \boldsymbol\tau \cdot \mathbf{b}_{p,E}(x)   ] \, + \, [ \boldsymbol\sigma \cdot \mathbf{b}_f(x) ] \otimes \tau_z \right\} \Psi
\nonumber \quad,
\end{eqnarray}
where $\boldsymbol\sigma=(\sigma_x,\sigma_y, \sigma_z)$ and $\boldsymbol\tau=(\tau_x,\tau_y, \tau_z)$ are two sets of Pauli matrices  acting on the spin space ($\sigma=\uparrow,\downarrow$), and on the chirality space ($\alpha=R,L$), respectively; $\sigma_0$ and $\tau_0$ are the $2 \times 2$ identity matrices in the related spaces, and  
\begin{equation}
\mathbf{b}_f(x)=\frac{\left(  |\Gamma_{f}(x)| \cos \phi_{f}(x)  ,  |\Gamma_{f}(x)| \sin \phi_{f}(x)  ,  eV_{f}(x) \right)}{ \hbar v_F}  \label{bf-def}
\end{equation}
\begin{eqnarray}
\lefteqn{\mathbf{b}_{p,E}(x)   = \frac{1}{\hbar v_F} \times }   & & \label{bpE-def} \\
& &  \times (-i\,  |\Gamma_{p}(x)|  \sin \phi_{p}(x)     , \,  i\, |\Gamma_{p}(x)|  \cos \phi_{p}(x)     \, , \,   eV_{p}(x) -E) \nonumber 
\end{eqnarray}
are two local `magnetic' fields,  determined by the tunnel junction parameter profiles, and acting on the spin and the chirality spaces, respectively. Finally, $E$ is the energy measured with respect to the Dirac point of the Hamiltonian (\ref{H0}). Notice that $\mathbf{b}_{p,E}$ and $\mathbf{b}_{f}$ enter the Hamiltonian (\ref{HAM-FER-MAT}) in a way that time-reversal symmetry is not broken.\cite{magnetic-fields} 

The stationary solutions $\Psi(x,t)=e^{-i E t/\hbar} \Psi_E(x)$ of the Heisenberg equation $i\hbar \, \partial_t  \Psi(x,t)=[ \Psi(x,t) \,, \hat{\mathcal{H}}]$ obtained from (\ref{HAM-FER-MAT}) fulfill the equation of motion
\begin{eqnarray}
 \lefteqn{i \,\frac{\partial }{\partial x}   \Psi_E(x)  
= } & &\label{eom-psi} \\
& & =  \left\{   [ \boldsymbol\sigma \cdot \mathbf{b}_f(x) ]\otimes \tau_0   \, +\sigma_0 \otimes [ \boldsymbol\tau \cdot \mathbf{b}_{p,E}(x)  ]  \,  \right\}  \Psi_E(x)  \quad, \nonumber 
\end{eqnarray} 
whose solution is factorized as a direct product
\begin{equation}
\label{sol}
\Psi_E(x)=\left(\mathbf{U}_f(x;0) \otimes \mathbf{U}_{p,E}(x;0) \right) \,  \Psi(0)\quad,  
\end{equation}
where the two `evolution' operators
\begin{eqnarray}
\mathbf{U}_f(x;0)&= &   {\rm T} \,\exp\left[{\displaystyle -i \int_{0}^x \, dx^\prime  \boldsymbol\sigma   \cdot \mathbf{b}_f(x^\prime)   }\right] \label{Uf-expr}\\
\mathbf{U}_{p,E}(x;0)&= &  {\rm T} \,\exp\left[{\displaystyle -i \int_{0}^x \, dx^\prime  \boldsymbol\tau  \cdot \mathbf{b}_{p,E}(x^\prime) \,} \right]\label{Up-expr}
\end{eqnarray}
are applied to the four-component field operator $\Psi(0)$ at the space origin. Here ${\rm T}$ denotes the space ordering, and plays a role similar to time ordering in time-dependent perturbation theory. As a consequence, the total transfer matrix of the junction also factorizes into a direct product
\begin{equation}
{\mathsf{M}}=  \mathbf{m}_f  \,   \otimes \, \mathbf{m}_p    \label{M-factorized} 
\end{equation} 
where
\begin{eqnarray}
\mathbf{m}_f & {\,=\,} & \mathbf{U}_f(x_f;x_0)  \label{mf-def}\\
\mathbf{m}_p & {\,=\,} &  e^{-i \tau_z k_E x_f}  \mathbf{U}_{p,E}(x_f;x_0)  e^{+i \tau_z k_E x_0} \, \, \, \label{mp-def}
\end{eqnarray}
determine the transmission coefficients (\ref{Tp-dep}) and (\ref{Tf-dep}), related to spin-preserving and spin-flipping processes, via the relations~\cite{sternativo}
\begin{eqnarray}
T_{p}&=&   |(\mathbf{m}_p)_{22}|^{-2}  \label{Tp-gen}\\
T_{f}&=&   |(\mathbf{m}_f)_{22}|^2 \label{Tf-gen}  \quad.
\end{eqnarray} 
We now notice that the random potentials $V_{p}(x)$ and $V_{f}(x)$ appear in the equations of motion (\ref{eom-psi}) through the $z$-components of the vectors $\mathbf{b}_f(x)$ and $\mathbf{b}_{p,E}(x)$ [see Eqs.(\ref{bf-def}) and (\ref{bpE-def})]. Thus, introducing the field
\begin{equation}
\Psi^\prime_E(x) {\,=\,} \left( \displaystyle e^{i \int_{x_0}^x  \frac{eV_{f}(x^\prime)}{\hbar v_F} \sigma_z} \otimes e^{i \int_{x_0}^x  \frac{eV_{p}(x^\prime)}{\hbar v_F}  \tau_z} \right) {\Psi}_E(x) \label{gauge-trans}
\end{equation}
it is straightforward to verify that, if $\Psi_E(x)$ fulfills Eq.(\ref{eom-psi}), then $\Psi^\prime_E$ fulfils a similar equation,
\begin{eqnarray}
 \lefteqn{ \, i \frac{\partial }{\partial x}  \Psi^\prime_E(x) =} &&  \label{eom-gauge}\\ & &  = \left\{ [\boldsymbol{\sigma}  \cdot \mathbf{b}^\prime_{f}(x) ] \otimes \tau_0 \, + \sigma_0 \otimes  [\boldsymbol{\tau}  \cdot \mathbf{b}^\prime_{p,E}(x) ] \,  \right\} \Psi^\prime_E(x)  \quad,\nonumber
\end{eqnarray}
with new vectors 
\begin{equation}
\mathbf{b}^\prime_f(x)=\frac{\left(  |\Gamma_{f}(x)| \cos \phi^\prime_{f}(x)  ,  |\Gamma_{f}(x)| \sin \phi^\prime_{f}(x)  ,  0 \right)}{ \hbar v_F}  \label{bf-tilde-def}
\end{equation}
\begin{eqnarray}
\lefteqn{\mathbf{b}^\prime_{p,E}(x)   =}   & & \label{bpE-tilde-def} \\
& & =\frac{(-i\,  |\Gamma_{p}(x)|  \sin \phi^\prime_{p}(x)     , \,  i\, |\Gamma_{p}(x)|  \cos \phi^\prime_{p}(x)     \, , \,   -E)}{\hbar v_F}\nonumber  \quad.
\end{eqnarray}
In Eqs.(\ref{bf-tilde-def}) and (\ref{bpE-tilde-def}) the potentials $V_{p/f}$ have disappeared from the $z$-components [see Eqs.(\ref{bf-def}) and (\ref{bpE-def}) for comparison], and have been absorbed into renormalized phases of the tunneling amplitudes, 
\begin{equation}
\left\{
\begin{array}{l}
\Gamma_{\nu}=|\Gamma_{\nu}| \, e^{i\phi_{\nu}} \, \\ \\
V_{\nu} \, \end{array}\right.  \rightarrow \hspace{1cm} \left\{
\begin{array}{l}  \Gamma^\prime_{\nu}=|\Gamma_{\nu}| \, e^{i\phi^\prime_{\nu}} \\ \\ V_{\nu} \equiv 0 \, \end{array}\right. \quad,\label{trans-prime}
\end{equation}
where $\phi^\prime_{\nu}(x)$ is given by Eq.(\ref{gauge-phase}).
For the $\Psi^\prime_E$-field the transfer matrices $\mathbf{m}^\prime_f$ and $\mathbf{m}^\prime_{p}$  are defined as in Eqs.(\ref{mf-def}) and (\ref{mp-def}), upon replacing $(\mathbf{U}_{f} \, , \, \mathbf{U}_{p,E})\rightarrow  (\mathbf{U}^\prime_{f} \, , \, \mathbf{U}^\prime_{p,E})$, where 
the latter   evolution operators  are defined as in Eqs.(\ref{Uf-expr}) and (\ref{Up-expr}) with $\mathbf{b}_{f}\rightarrow \mathbf{b}^\prime_{f}$ and $\mathbf{b}_{p,E}\rightarrow \mathbf{b}^\prime_{p,E}$. From Eq.(\ref{gauge-trans}) one can easily show that
\begin{eqnarray}
\mathbf{U}^\prime_{f}(x_f;x_0) &= & e^{i \int_{x_0}^{x_f} \frac{eV_{f}(x^\prime)}{\hbar v_F}\, \sigma_z} \, \mathbf{U}_{f}(x_f;x_0) \label{Uftilde-expr} \\
& & \nonumber \\
\mathbf{U}^\prime_{p,E}(x_f;x_0) &= &   e^{i \int_{x_0}^{x_f} \frac{eV_{p}(x^\prime)}{\hbar v_F} \, \tau_z} \, \mathbf{U}_{p,E}(x_f;x_0) \, \quad, \label{Uptilde-expr} 
\end{eqnarray}
and that the transmission coefficients are independent of the gauge transformation, as expected. \\

%%%%%%%%%%%%%%%%%%%%%%%%%%%%%%%%%%%%%%%%%%%%%%%%%
%%%%%%%%%%%%%%%%%%%%%%%%%%%%%%%%%%%%%%%%%%%%%%%%% 
%%%%%%%    s i n g l e   s a m p l e    %%%%%%%%%
%%%%%%%%%%%%%%%%%%%%%%%%%%%%%%%%%%%%%%%%%%%%%%%%%
%%%%%%%%%%%%%%%%%%%%%%%%%%%%%%%%%%%%%%%%%%%%%%%%%
\section{Single samples}
\label{sec-III}
\subsection{Introduction}
The factorisation result (\ref{M-factorized}), valid for any arbitrary profile of the tunneling amplitudes and potentials, implies the separated dependence of the two coefficients $T_p$ and $T_f$ on the disorder profiles $\Gamma_p(x),V_{p}(x)$ and  $\Gamma_f(x),V_{f}(x)$, respectively [see Eqs.(\ref{Tp-dep}) and (\ref{Tf-dep})]. This enables us to analyze the effects of disorder in the two sectors $\nu=p,f$ separately,   without loss of generality.  In particular,   because the spin-preserving tunneling amplitude is typically bigger then the spin-flipping one, $|\Gamma_p|>|\Gamma_f|$, and because $T_p$ exhibits a richer energy-dependence than $T_f$,~\cite{sternativo} we shall consider henceforth the {p}-sector, for which the spin index acts as a dummy degeneracy variable.  Furthermore, because the potential $V_{p}$ can be reabsorbed via the gauge transformation (\ref{trans-prime}) into the phase of the tunneling amplitude [see Eq.(\ref{gauge-phase})], we shall discuss without loss of generality the effects of disorder of the (complex) tunneling amplitude $\Gamma_p$.  

We start our analysis by investigating the transmission coefficient of individual disordered samples. 
To illustrate the effects of disorder, it is first worth recalling the result for the clean junction case, characterised by a constant $\Gamma_p(x) \equiv|\Gamma_p| \exp[i \phi_p]$ over the whole junction length~$L$, as investigated in Ref.[\onlinecite{sternativo}].
In that case the transmission coefficient $T_p$ is {\it independent} of the value of the phase~$\phi_p$, and depends on only the absolute value $\Gamma_p^0 \!:= |\Gamma_p|$ of the tunneling amplitude. In particular, while for a short junction ($L<\xi_0$ with $\xi_0 \!:= \hbar v_F/\Gamma_p^0$) the transmission coefficient is trivially energy independent,  for an elongated junction ($L>\xi_0$)  the energy $\Gamma_p^0$ determines the crossover from a `sub-gap' region ($E<\Gamma_p^0$) with very low transmission to a `supra-gap'  region ($E>\Gamma_p^0$) where transmission is finite and exhibits an oscillatory behavior related to the finite length $L$ of the junction.  In the limit of an infinitely long junction, $\Gamma_p^0$   would be an actual gap in the spectrum, so that  $\xi_0$ is the evanescent wave decay length associated with such an energy gap.  We   thus focus here on the effects of disorder in the more interesting case of an elongated junction $L>\xi_0$.

For this purpose, we perform a coarse graining of the tunnel junction, dividing the length $L$ of the tunnel region into a sequence of $N$ intervals $[x_{j-1}; x_j]$ (with $j=1,\ldots N$ and $x_N=x_f$), characterized by a size $l_p=L/N$ corresponding to the typical disorder fluctuation lengthscale.   Then, in each interval $j$ we generate local model parameters according to specific distributions (see below), thereby determining its local transfer matrix $\mathbf{m}_p^{(j)}$. The transfer matrix $\mathbf{m}_p$ of the whole sample is then  obtained as the product of the transfer matrices in each interval,~\cite{ihn-book} 
\begin{equation}
\mathbf{m}_p = \prod_{j=N}^1 \mathbf{m}_p^{(j)} \quad. \label{m-as-a-prod}
\end{equation}
The $\mathbf{m}_p^{(j)}$'s of each interval can be computed exactly  for the case of locally constant tunneling amplitude, and also for the case of a phase $\phi_p$ varying linearly within the interval  (see the Appendix).  
The transmission coefficient~$T_p$ of the disordered sample, obtained from (\ref{m-as-a-prod}) through Eq.(\ref{Tp-gen}), depends both on the fluctuations strength of the tunneling amplitude and on  the lengthscales $L$, $\xi_0$ and $l_p$. Here we shall focus on the regime $l_p < \xi_0 < L$, which is physically more relevant, as we shall discuss later.

%%%%%%%%%%%%%%%%%%%%%%%%%%%%%%%%%%
%%%%%%%%%%%%%%%%%%%%%%%%%%%%%%%%%%
%%%%%%%%%%%%%%%%%%%%%%%%%%%%%%%%%%
\subsection{Roles  of the absolute value and phase of the tunneling amplitude with disorder}
In the presence of disorder, the roles of the absolute value $|\Gamma_p(x)|$ and of the phase $\phi_p(x)$ can be quite different from those for the clean junction case. In particular,    while  in the clean case $T_p$ is independent of the phase, in the disordered case the fluctuations of $\phi_p(x)$ dramatically affect~$T_p$.  

\begin{figure} 
\centering
\includegraphics[width=\columnwidth,clip]{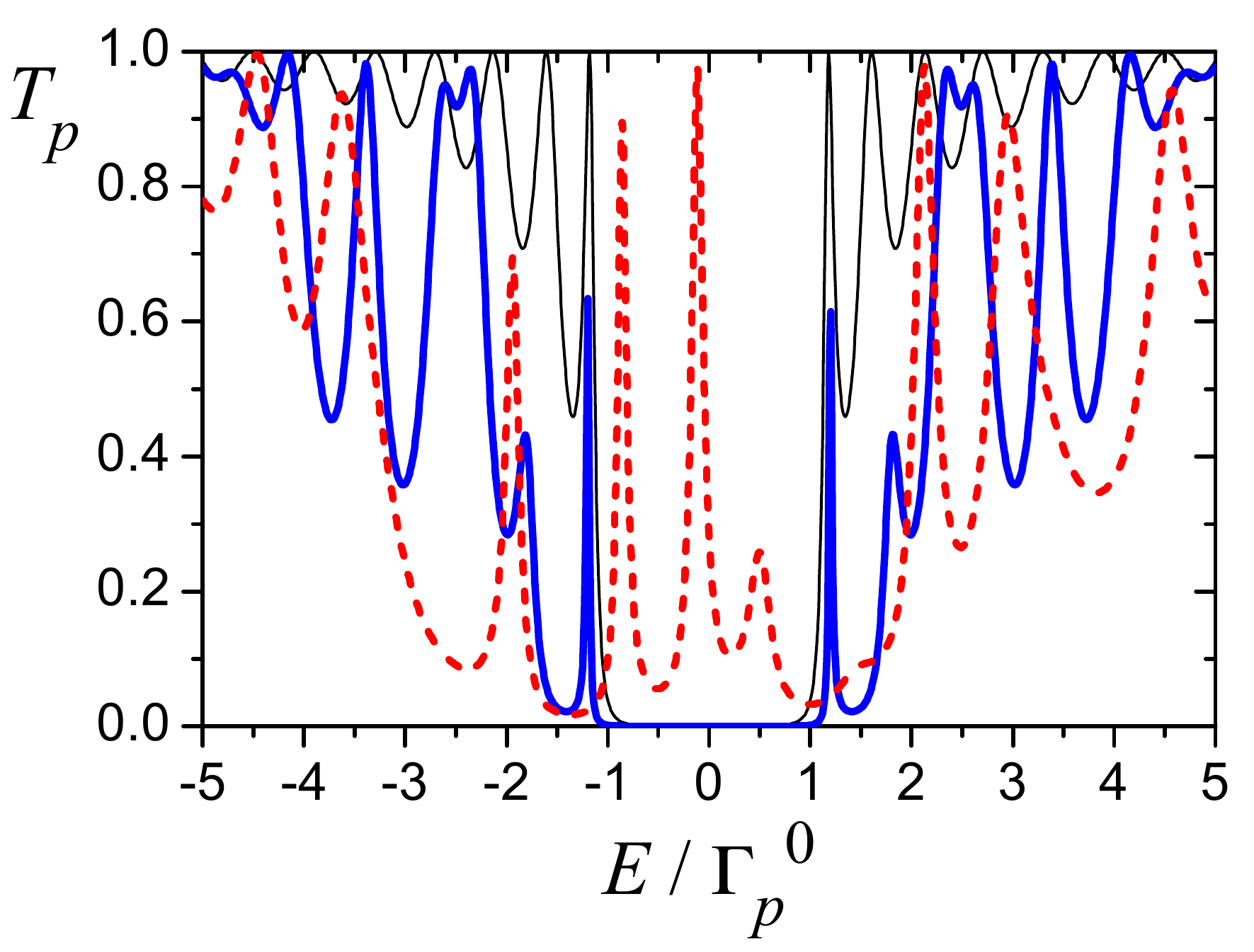}
\caption{(Color online) Transmission coefficient for a disordered tunnel junction where the (complex) tunneling amplitude $\Gamma_p=|\Gamma_p| \exp[i\phi_p]$ is randomly varying. The disorder on the absolute value $|\Gamma_p|$ (solid blue curve) has different effects than the disorder on the phase $\phi_p$ (dashed red curve). For comparison, the thin black line denotes the clean case.  Here $L/\xi_0=5$ and $N=10$, corresponding to the regime $l_p  \lesssim \xi_0 < L$.}
\label{Fig-SGL-Gamma_1}
\end{figure}

We illustrate this point with an illuminating example, namely, the case where $\Gamma_p$ in each interval $j$ are   chosen  from a Gaussian distribution in the complex plane. Then, the absolute value $|\Gamma_p|$  follows the Rayleigh distribution and the phase $\phi_p$ is uniformly distributed between 0 and $2\pi$. We have analyzed independently the effects of these two distributions on the transmission coefficient $T_p$. Explicitly, Fig.\ref{Fig-SGL-Gamma_1} shows the $T_p$ of a disordered sample in the cases where (i) $|\Gamma_p|$ is Rayleigh-distributed and $\phi_p$ is kept constant (solid blue curve), and (ii) $\phi_p(x)$ fluctuates uniformly and $|\Gamma_p|$ is constant (dashed red curve). For comparison, the case of a clean junction is described by the thin black curve, and the Rayleigh distribution is taken with an average value $\langle |\Gamma_p| \rangle=\Gamma_p^0$ equal to the absolute value $\Gamma_p^0$ of the clean case. As one can see, the fluctuations of the absolute value $|\Gamma_p|$ leave essentially unaltered the qualitative features of the clean case. Indeed, the particle-hole symmetry $T_p(E)=T_p(-E)$ of the transmission coefficient is preserved, and the crossover between a low transmission `sub-gap' region and a high transmission `supra-gap' region is still present, possibly with a slight quantitative change of the `gap' $|\Gamma_p|$ and a modification of the $T_p$ oscillations in the supra-gap region.  In contrast, the disorder of the phase $\phi_p$ breaks the particle-hole symmetry of the clean case, and  introduces transmission peaks in the `sub-gap' region, whose width decreases when $L/\xi_0$ increases. 
The origin of these peaks can be qualitatively understood in analogy to the case of barriers in quantum interference problems. In the simplest case of a tunneling amplitude with a phase jump in the middle of the junction, for instance, the two halves of the junction play the role of a series of two barriers, located very close to each other, with equal transmission coefficients (here determined by $|\Gamma_p|L/2$), but with a difference in the transmission amplitude phases (here determined by $\Delta \phi_p$ and $E$): perfect resonances are then known to arise even when each individual barrier is not perfectly transmitting, provided that the phase difference fulfills specific conditions (here $\Delta \phi_p=\pi$ and $E=0$). In a more general case where phase fluctuations occur at various junction positions and are not necessarily equal to $\pi$, as in Fig.\ref{Fig-SGL-Gamma_1}, peaks do not reach exactly 1 and may occur at different values of energy $E$.\\
The phase disorder thus significantly  modifies the clean case scenario, replacing the crossover between the two regions with a sequence of  maxima and minima, and breaking the particle-hole symmetry. 
%%%%%%%%%%%%%%%%%%%%%%%%%%%%%%%%%%
%%%%%%%%%%%%%%%%%%%%%%%%%%%%%%%%%%
While the specific shape of $T_p$ is sample-dependent, the qualitative features described above in Fig.\ref{Fig-SGL-Gamma_1}  are present in any sample in the same regime of parameters, namely $l_p \lesssim \xi_0 < L$.

\begin{figure} 
\centering
\includegraphics[width=\columnwidth,clip]{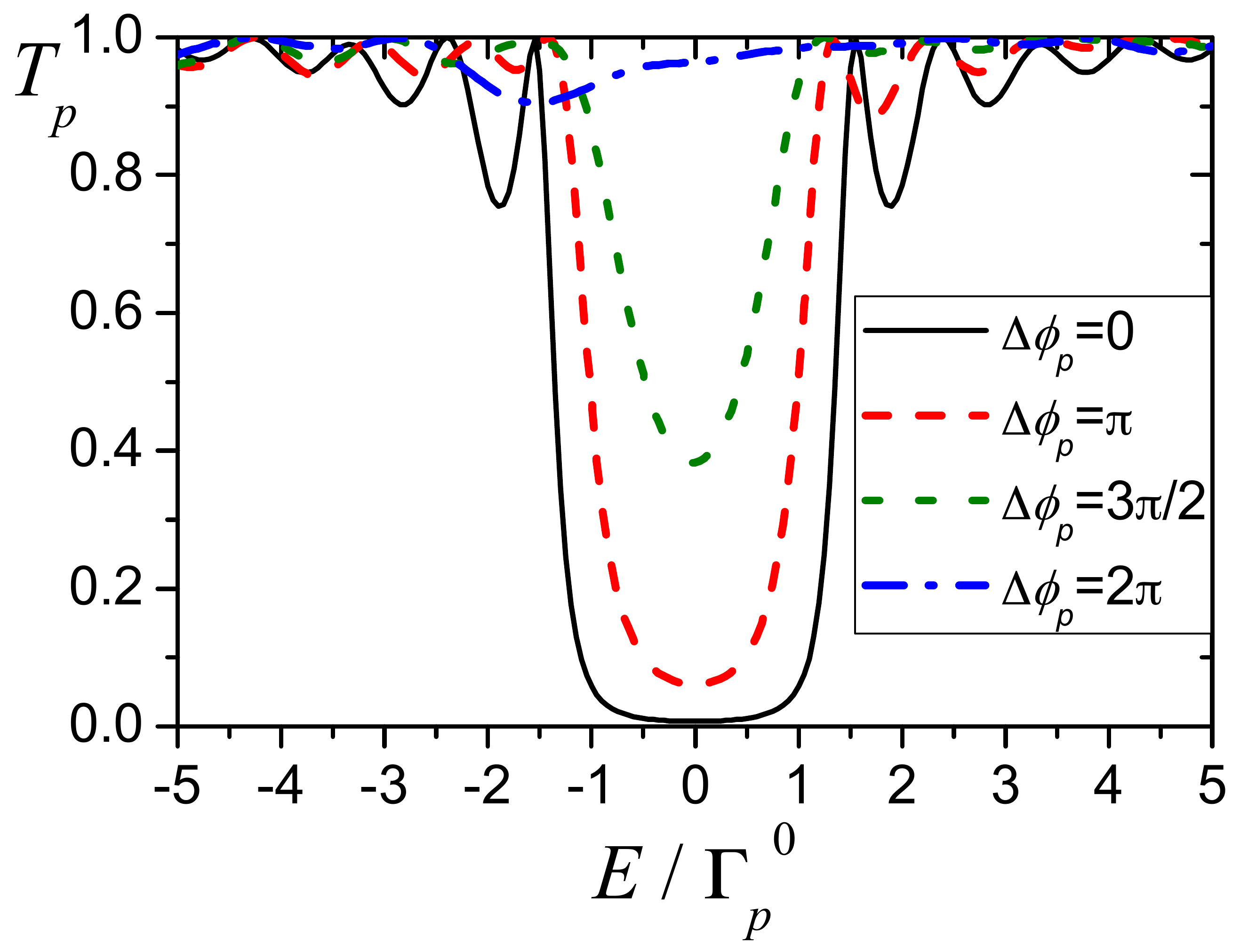}
\caption{(Color online) Transmission coefficient for a disordered tunnel junction with $L/\xi_0=3$ and $N=500$, corresponding to the regime $l_p \ll \xi_0 <L$. The absolute value $|\Gamma_p|$ varies according to the Rayleigh distribution, and the phase is uniformly picked in a range $[-\Delta \phi_p/2; \Delta \phi_p/2]$. While the fluctuations of the absolute value $|\Gamma_p|$ of the tunneling amplitude do not modify the clean case result significantly, the amount $\Delta \phi_p$ of  the phase  fluctuations may have a dramatic impact. In particular, while for $\Delta \phi_p < \pi$ minor modifications appear on $T_p$, when $\Delta \phi_p \simeq 2\pi$ (corresponding to a Gaussian distributed tunneling amplitude) the phase fluctuations wash out the tunneling term, so that the transmission coefficient becomes energy independent and tends to 1. }
\label{Fig-SGL-Gamma_2}
\end{figure}

The difference in the role of $|\Gamma_p|$ and $\phi_p$ becomes particularly striking in the regime $l_p \ll \xi_0 < L$, where disorder  fluctuates over a length scale much smaller than the  typical electronic scale characterizing the clean junction at $E=0$. In this limit the integral in (\ref{Hsp}) exhibits a slowly varying component [the electron field $\Psi(x)$] and a rapidly fluctuating  term [the tunneling amplitude $\Gamma_p(x)$], and the electron probes only the space-averaged disorder $\langle  \Gamma_p  \rangle_x$, which is a statistical estimate of the distribution average itself. Thus, when only $|\Gamma_p|$ fluctuates, one has $\langle  \Gamma_p  \rangle_x \simeq \langle |\Gamma_p| \rangle \exp[i\phi_p]=\Gamma_p^0 \exp[i\phi_p]$ and the clean case result is recovered. In contrast, if $\phi_p$ fluctuates of about $2\pi$, then for any value of $\Gamma_p$, also $-\Gamma_p$ is equally likely to occur, so that $\langle  \Gamma_p  \rangle_x \simeq 0$. The phase fluctuations completely wash out the tunneling term, so that the transmission becomes energy independent and tends to~1. This is illustrated in Fig.\ref{Fig-SGL-Gamma_2}, which displays the transmission coefficient for a sample where both $|\Gamma_p|$ and $\phi_p$ are disordered over a length $l_p=L/500$. In particular, the former is Rayleigh-distributed and the latter is uniformly sampled within a range $\Delta \phi_p$. As one can see, when $\Delta \phi_p \le \pi$, the transmission coefficient does not significantly change with respect to the clean case (see thin black curve in Fig.\ref{Fig-SGL-Gamma_1} for comparison), whereas when $\Delta \phi_p $ approaches $2\pi$ the transmission coefficient rapidly rises towards 1,  becoming independent of $E$. \\

%%%%%%%%%%%%%%%%%%%%%%%%%%%%%%%%%%
%%%%%%%%%%%%%%%%%%%%%%%%%%%%%%%%%%

These results about  single samples indicate that, already with the customary assumption of a Gaussian distributed tunneling amplitude $\Gamma_p$  (i.e. a Rayleigh distributed  $|\Gamma_p|$ and a uniformly distributed $\phi_p$), the fluctuations of the  phase  have a much more significant impact on $T_p$ than those of the absolute value. Also,  as we shall discuss in Sec.\ref{sec-V}, the Rayleigh distribution used in Figs.\ref{Fig-SGL-Gamma_1} and \ref{Fig-SGL-Gamma_2}  in fact already greatly overestimates the actual amount of fluctuations of $|\Gamma_p|$ of realistic systems. For these reasons, we shall henceforth neglect the disorder on $|\Gamma_p|$ and focus on the phase. In the next section we show that different models for the phase fluctuations may lead to strongly different results also for the localization length of disordered tunnel junctions.

%%%%%%%%%%%%%%%%%%%%%%%%%%%%%%%%%%%%%%%%%%%%%%%%%%%%%%%
%%%%%%%%%%%%%%%%%%%%%%%%%%%%%%%%%%%%%%%%%%%%%%%%%%%%%%%
%%%%%%  L o c a l i s a t i o n   l e n g t h  %%%%%%%%
%%%%%%%%%%%%%%%%%%%%%%%%%%%%%%%%%%%%%%%%%%%%%%%%%%%%%%%
%%%%%%%%%%%%%%%%%%%%%%%%%%%%%%%%%%%%%%%%%%%%%%%%%%%%%%%
\section{Localization length and sample averaging}
\label{sec-IV}
This section is devoted to the investigation of intrinsic disorder effects on helical tunnel junctions, i.e. effects that do not depend on the specific sample.  
The localization length $\xi_{loc}$ is an intrinsic property of disorder that determines the exponential decay of the transmission coefficient of a disordered sample with its length $L$, 
\begin{equation}
T_p(E) \sim e^{-2 L /\xi_{loc}(E)} \quad.
\end{equation}
It can be computed as a Lyapunov exponent~\cite{kramer-review,vulpiani_book,delande-review}: Denoting by $\lambda^{max}(M)$ the maximal eigenvalue of the transfer matrix for a system with $M$ disordered intervals, one has
\begin{equation}
\xi_{loc}^{-1}    {\,:=\,} \frac{1}{l_p} \lim_{M \rightarrow \infty} \frac{\ln |\lambda^{max}(M)|}{M}
\quad, 
\end{equation}
where $l_p$ is the length of each interval. In practice, a convergence is reached for values of $M \sim 10^3$, and one can see that $\xi_{loc}$ is sample independent, in agreement with Oseledec's theorem.\cite{oseledec,vulpiani_book}\\

To identify how disorder affects $\xi_{loc}$, it is first necessary to specify the length dependence of the transmission in the clean case, which is not necessarily a conducting system. Indeed, from the expression for $T_p$ for a clean junction,\cite{sternativo}
\begin{eqnarray}
\lefteqn{T_{p}(E) = \hspace{2cm}} &   \label{Tp-const}  \\
& & =\left\{ \begin{array}{lll}  \displaystyle \left[1+\frac{{\Gamma_{p}^0}^2}{E^2-{\Gamma_{p}^0}^2}\, \sin^2{\left(\frac{L}{\xi_0}\sqrt{\frac{E^2}{{\Gamma_{p}^0}^2}-1}  \,  \right) } \right]^{-1} &   |E|> \Gamma_{p}^0
\\ & & \label{Tp-theta} \\
\displaystyle \left[1+\frac{{\Gamma_{p}^0}^2}{{\Gamma_{p}^0}^2-E^2}\, \sinh^2{\left(\frac{L}{\xi_0}\sqrt{1-\frac{E^2}{{\Gamma_{p}^0}^2}}  \,  \right) } \right]^{-1} &  |E|<\Gamma_{p}^0
\end{array} \right.  \nonumber
\end{eqnarray}
one can straightforwardly read off the `localization length' for the clean case 
\begin{equation}
\xi_{cl}(E)= \left\{ 
\begin{array}{lcl} 
\xi_0 \left( 1-(E/\Gamma_{p}^0)^2 \right)^{-1/2} & \mbox{for} & |E|<  \Gamma_{p}^0  \\ & & \\
\infty  & \mbox{for} & |E|>  \Gamma_{p}^0 
\end{array}
\right. \label{xi-cl}
\end{equation}
where $\xi_0=\hbar v_F/\Gamma_p^0$ is the decay length of the clean case wavefunction at $E=0$.
Below  we show how the behavior (\ref{xi-cl}) is modified in the presence of disorder.

%%%%%%%%%%%%%%%%%%%%%%%%%%%%%%%%%%%%%%%%%%%%%%%%%%%%%%%%%%%%%%%%%%%%%%%%
%%%%%%%%%%%%%%%%%%%%%%%%%%%%%%%%%%%%%%%%%%%%%%%%%%%%%%%%%%%%%%%%%%%%%%%%
%%%%%%%%%%%%%%%%%%%%%%%%%%%%%%%%%%%%%%%%%%%%%%%%%%%%%%%%%%%%%%%%%%%%%%%%

\subsection{Localization length: Comparison between three different models for phase fluctuations}
\label{Sec-IV-a}
\begin{figure} 
\centering
\includegraphics[width=8cm,clip]{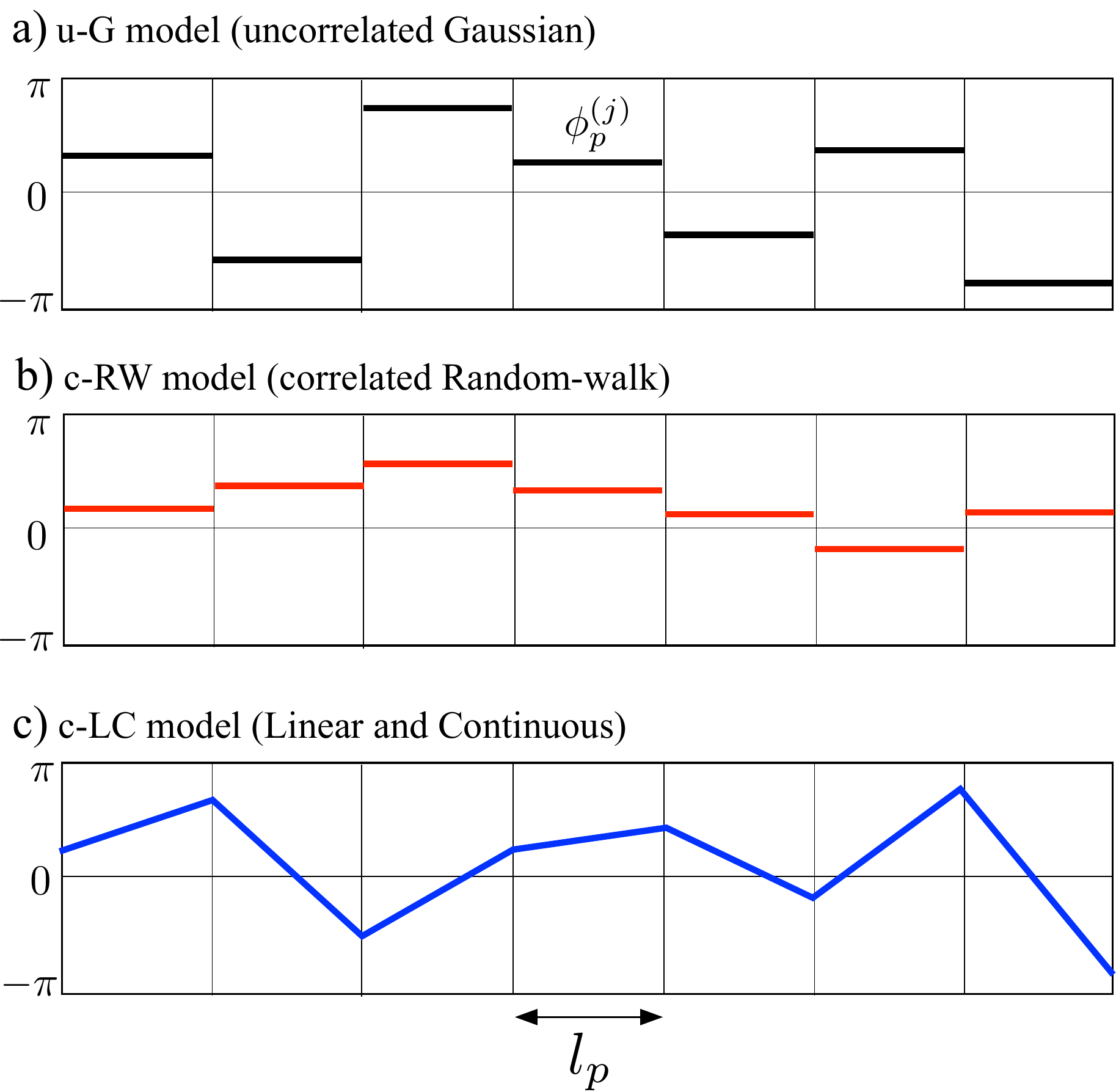}
\caption{(Color online) Three different models of disorder fluctuations for the phase $\phi_p$ of the tunneling amplitude [see Eqs.(\ref{phase-u-G})-(\ref{phase-c-RW}) and (\ref{phase-c-LC})]. The horizontal axis represent the longitudinal coordinate along the junction, divided into $N$ intervals with a size $l_p$.}
\label{Fig-phase-models}
\end{figure}

The results of Sec.\ref{sec-III} obtained for single samples suggest that, in the regime $l_p < \xi_0 < L$, the fluctuations of the phase of the tunneling amplitude matter more than those of its absolute value. 
Here we analyze three models for phase fluctuations, and show that they lead to quite different predictions for the localization length of  disordered tunnel junctions. In Sec.\ref{sec-V} we shall  discuss which model better suits which situation. The three models differ in the way the phase $\phi_p$ is assumed to vary from one interval to another, and are schematically depicted in Fig.\ref{Fig-phase-models}: \\

\noindent a)  {\it Uncorrelated Gaussian (u-G) Model}. The phase $\phi_p$ is generated independently from one interval to another, according to a uniform distribution `{\rm  ran}' between $-\pi$ and $\pi$ [see Fig.\ref{Fig-phase-models}a)]
\begin{equation}
\phi_p^{(j)}={\rm ran}(-\pi;+\pi) \label{phase-u-G}
\end{equation}
The name `Gaussian' originates from the fact that, when the values of a complex tunneling amplitudes $\Gamma_p=|\Gamma_p|\exp[i \phi_p]$ are generated from a  Gaussian distribution, Eq.(\ref{phase-u-G}) is the distribution for their phases;\\

\noindent b) {\it correlated  random-walk  (c-RW) model}. The phase $\phi_p^{(j+1)}$ in the $(j+1)$-th interval deviates only by an amount $\Delta \phi_p$ from the phase in $\phi_p^{(j)}$ in the $j$-th interval, namely it is generated as
\begin{equation}
\phi_p^{(j+1)}=\phi_p^{(j)}+ {\rm ran}(-\frac{\Delta\phi_p}{2};+\frac{\Delta\phi_p}{2})  \label{phase-c-RW}
\end{equation}
The phase randomly evolves by `steps' of (at most) $\pm \Delta\phi_p/2$  to the `right'/`left' with respect to the value in the previous interval, similarly to a  random-walk [see Fig.\ref{Fig-phase-models}b)]. When the `step' parameter $\Delta\phi_p \rightarrow 2\pi$, the c-RW Model reduces to the u-G Model;\\

\noindent c)  {\it Correlated linear and continuous (c-LC) model}. The phase $\phi_p$ is assumed to vary linearly within each interval $l_p$, with a random slope $-2k_p^{(j)}$. However, the phase is assumed to be continuous, so that the extremal values of $\phi_p$ in the $j$-th interval have to match the ones of the neighboring intervals $j\pm1$~[see Fig.\ref{Fig-phase-models}c)]. This amounts to setting
\begin{eqnarray}
\phi_p^{(j)}(x) &=& -2k_p^{(j)} (x-x_{j-1}) \,-2  l_p \sum_{i=1}^{j-1} k_p^{(i)} \hspace{0.4cm} 
\label{phase-c-LC} \\
& & \hspace{4cm} \mbox{for } x_{j-1} \le x \le x_{j} \, .\nonumber
\end{eqnarray}
We shall assume that the $k_p^{(j)}$'s are generated from a Gaussian distribution, with a vanishing average and with  a standard deviation ${\rm SD}(k_p)$, which is the parameter characterizing this model. When $2l_p \, {\rm SD}(k_p)  \ll 2 \pi$, the clean case is recovered.\\

%%%%%%%%%%%%%%%%%%%%%%%%%%%%%%%%%%
%%%%%%%%%%%%%%%%%%%%%%%%%%%%%%%%%%

\begin{figure} 
\centering
\includegraphics[width=8cm,clip]{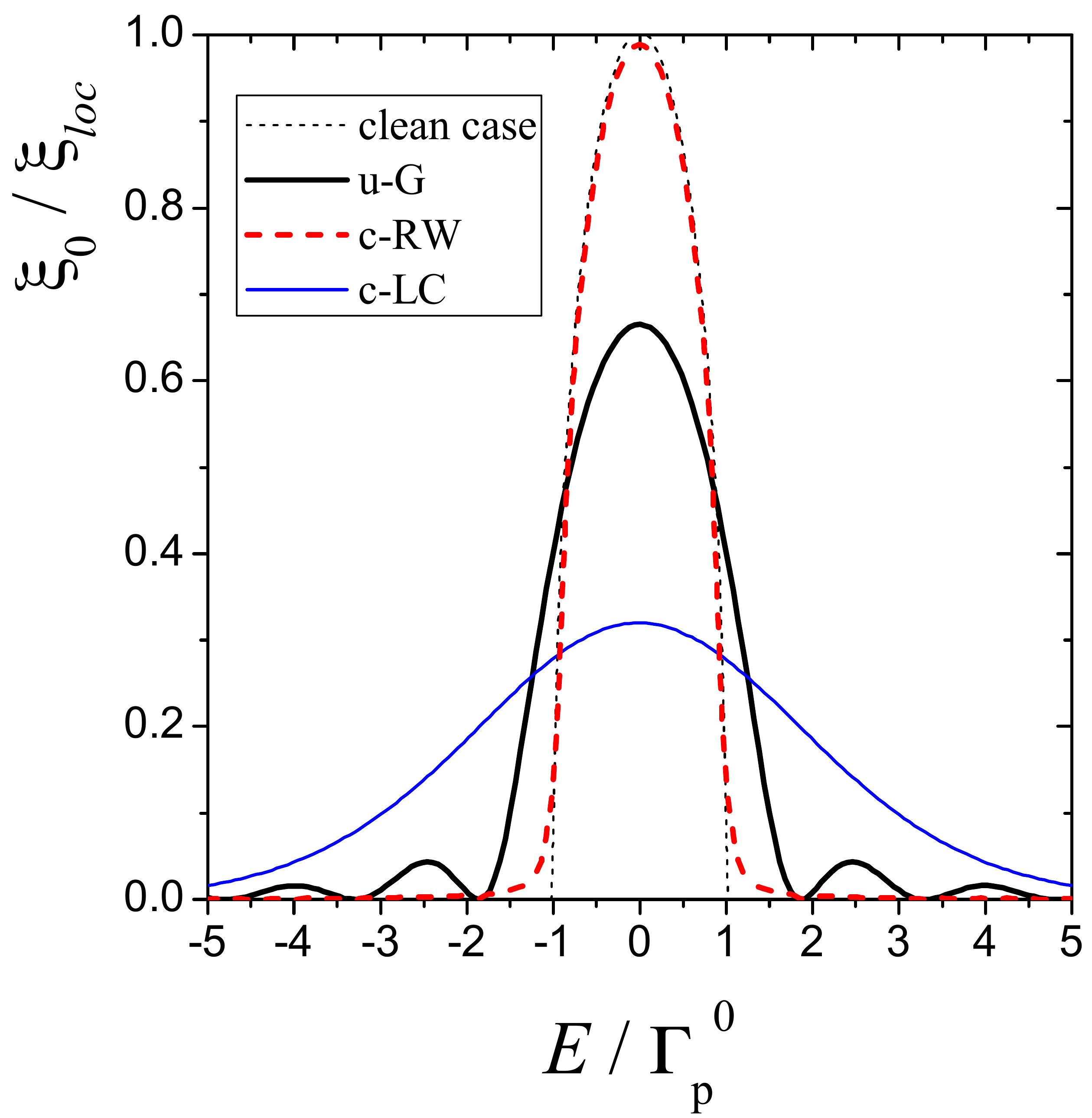}
\caption{(Color online) The (inverse) localization length $\xi_{loc}$, in units of $\xi_0$, is plotted as a function of energy for the three different models of phase fluctuations depicted in Fig.\ref{Fig-phase-models} and described in the text. The quite different behaviors are discussed in the text. The disorder length scale is $l_p=2\xi_0$ for all models. For the c-RW model $\Delta\phi_p=\pi/2$, and for  the c-LC model the parameter is 
${\rm SD}(k_p)=2/\xi_0$. The dotted black curve represents the clean case result $\xi_0/\xi_{cl}(E)$ [see Eq.(\ref{xi-cl})]. }
\label{Fig-xi-loc}
\end{figure}

We have computed the localization length as a function of the energy $E$ for these three models, obtaining quite different results, as shown in Fig.\ref{Fig-xi-loc}. All the curves refer to an interval length $l_p = 2\xi_0$. For comparison, the `localization length' $\xi_{cl}(E)$ of the clean case  [see Eq.(\ref{xi-cl})] is plotted with a dotted black curve, and exhibits a single lobe within the gap. 

For the u-G model (black solid curve), $\xi_{loc}$ has a non-monotonous energy dependence: this type of disorder transforms the single-lobe of the clean case  into a `Fraunhofer-like' pattern of lobes, with decreasingly  high maxima, which extend also in the `supra-gap' region and are separated by energy values of infinite localization length. These energy values correspond to resonances related to the lengthscale $l_p$ of the disorder. Indeed, although in each interval $j$ the phase fluctuates randomly, for the particular energy values $E=\pm \sqrt{{\Gamma_p^0}^2+(m \pi \hbar v_F/l_p)^2}$ ($m=1,2,\ldots$) the transfer matrix $\mathbf{m}_p^{(j)}$ becomes proportional to the identity matrix, independent of the local random value $\phi_p^{(j)}$ of the phase (see the Appendix).

In contrast, the c-RW model (red dashed curve) leaves the clean case behavior essentially unchanged, with the only visible effect of smoothening the transition at $E=\Gamma_p^0$ from the `sub-gap' to the `supra-gap' region, where the localization length is reduced to a finite value with respect to the clean case. Notice that the phase-step parameter $\Delta \phi_p$ of the c-RW model [see Eq.(\ref{phase-c-RW})] has been purposely chosen to be smaller but not much smaller than $2\pi$ ($\Delta \phi_p=\pi/2$), to emphasize that the result of the clean case is rather robust to such type of phase disorder, unless $\Delta \phi_p$ really approaches $2\pi$.   

Finally, the c-LC model of phase fluctuations (blue thin solid curve) transforms the sharp lobe of the clean case into a smooth Gaussian-like energy profile, with a standard deviation roughly given by $\hbar v_F {\rm SD}(k_p)$. Any signature of  crossover between a `sub-gap' region and a `supra-gap' region has disappeared. In comparison with the clean case (dotted black curve), the c-LC disorder lowers the maximum and broadens the curve; that is, it increases the localization length at low energies, and  reduces it for high energies. Here the model parameter has been chosen to be ${\rm SD}(k_p) =2/\xi_0$, so that across each interval $l_p$ the phase may change up to $\sim 2\pi$. \\

%%%%%%%%%%%%%%%%%%%%%%%%%%%%%%%%%%
%%%%%%%%%%%%%%%%%%%%%%%%%%%%%%%%%%

The comparison between the results of these three models indicates that the localization length of the clean junction is robust to disorder of phase fluctuations, unless the their typical amount $\Delta \phi_p$ is {\it significantly} close to~$2\pi$. Indeed one can show that, when $\Delta \phi_p\ll  2\pi$,  the three models lead to quite similar results, which do not qualitatively deviate from the clean case. In contrast, when $\Delta \phi_p \sim 2\pi$, the {\it way} the phase varies from one interval to another does matter in determining $\xi_{loc}$, as is appearent from the different behaviors of the u-G model (black solid curve) and the c-LC model (blue thin solid curve) in Fig.\ref{Fig-xi-loc}. We observe that, while both models deviate from the clean case, at a given value of $\xi_0/L$ the c-LC model exhibits the most striking difference. Indeed, the effect of a linearly varying phase is to produce a local shift of the electron energy $E\rightarrow E-\hbar v_F k_p^{(j)}$, related to the phase variation rate $-2k_p^{(j)}$ in Eq.(\ref{phase-c-LC}). Such an energy shift does not occur for   piecewise constant phase fluctuations, as can be seen by comparing the expressions for the electron field evolution operator $\mathbf{U}_{p,E}$  within a given interval $j$
in the two cases, explicitly given in the Appendix, Eqs.(\ref{U-c-uG_1}) and (\ref{U-c-uG_2}) and (\ref{U-c-LC_1}) and (\ref{U-c-LC_2}). Because the transmission coefficient is strongly energy dependent, such an energy shift can effectively displace the local transmission of that interval from the `sub-gap' regime to the `supra-gap' regime and viceversa, even when $|\Gamma_p|$ is not fluctuating. When the sequence of all intervals is now considered, for a value of energy $E$ corresponding to the `sub-gap' (`supra-gap') region of the clean case, a c-LC disordered sample also exhibits intervals in the conducting `supra-gap' (`sub-gap') regime. This is the reason why for the blue thin solid curve $\xi_{loc}(E)$ increases (decreases) for $|E|< \Gamma_p^0$ ($|E|> \Gamma_p^0$), as compared to the thin dotted black curve for the clean case. At a more formal level, this effect stems from the fact that a linearly varying phase $\phi_p$ results into a space-dependent $\mathbf{b}_{p,E}(x)$ vector (\ref{bpE-def}), so that the matrices $\boldsymbol\tau \cdot \mathbf{b}_{p,E}(x)$ and $\boldsymbol\tau \cdot \mathbf{b}_{p,E}(x^\prime)$ at any two different points do not commute, even within each interval. Space-ordering ${\rm T}$ in Eq.(\ref{Up-expr}) is thus crucial in determining the evolution operator $\mathbf{U}_{p,E}$, and gives rise to the local  energy shift $E\rightarrow E-\hbar v_F k_p^{(j)}$. In contrast, for piece-wise constant $\phi_p(x)$ and $\boldsymbol\tau \cdot \mathbf{b}_{p,E}(x)$, space-ordering can be dropped within each interval, and no shift arises.  
%%%%%%%%%%%%%%%%%%%%%%%%%%%%%%%%%%%%%%%%%%%%%%%%%%%%%%%%%%%%%%%%%%%%
%%%%%%%%%%%%%%%%%%%%%%%%%%%%%%%%%%%%%%%%%%%%%%%%%%%%%%%%%%%%%%%%%%%%
%%%%%%%%%%%%%%%%%%%%%%%%%%%%%%%%%%%%%%%%%%%%%%%%%%%%%%%%%%%%%%%%%%%%
%%%%%%%%%%%%%%%%%%%%%%%%%%%%%%%%%%%%%%%%%%%%%%%%%%%%%%%%%%%%%%%%%%%%
%%%%%%%%%%%%%%%%%%%%%%%%%%%%%%%%%%%%%%%%%%%%%%%%%%%%%%%%%%%%%%%%%%%%
\subsection{Sample averaging}
\begin{figure} 
\centering
\includegraphics[width=8cm,clip]{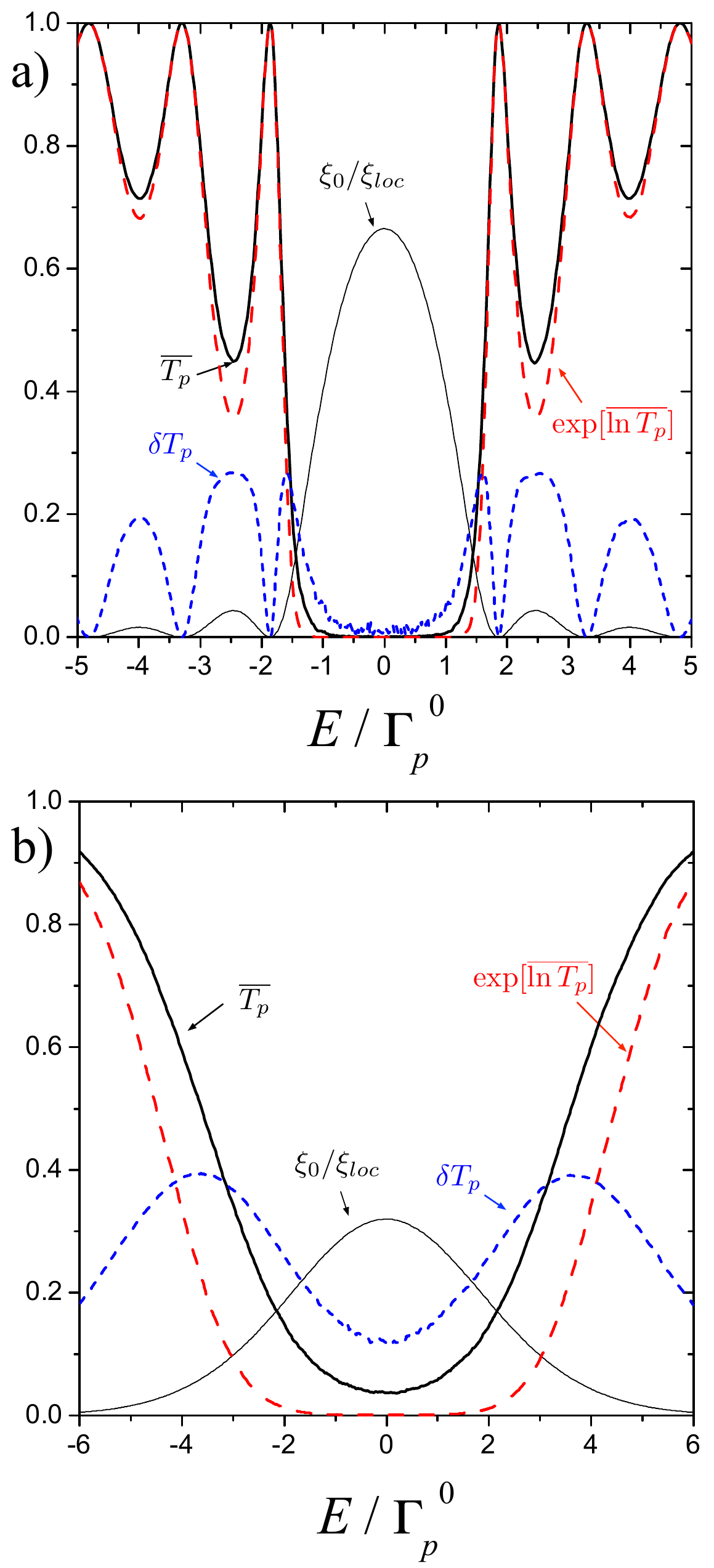}
\caption{(Color online) Energy dependence of sample-averaged transmission: the average transmission $\overline{T_p}$ (black solid curve), the `typical' transmission $T_p^{typ}$ [see Eq.(\ref{Tp-typ-def})] (red dashed curve), and the transmission fluctuations $\delta T_p$ [see (\ref{deltaTp})] (blue dotted curve). Here $l_p = 2\xi_0$, $N=6$ and averaging has been performed over $6 \times 10^3$ samples. a) the u-G model and b) the c-LC model for phase fluctuations (see Fig.\ref{Fig-phase-models}). In both cases the behavior of $\xi_0/\xi_{loc}$ has also been reported (thin black solid curve) as a guide to the eye. The disorder parameter for the c-LC model is ${\rm SD}(k_p)=2/\xi_0$.}
\label{Fig-SMP-AVE}
\end{figure}

The energy dependence of the localization length directly impacts sample-averaged quantities. Denoting by $\overline{(\ldots)}$ the average over samples with different disorder realizations, the sample-averaged transmission coefficient~$\overline{T_p}$, the `typical' transmission coefficient 
\begin{equation}
T_p^{typ} {\,=\,}\exp[\overline{\ln T_p}] \quad, \label{Tp-typ-def}
\end{equation}
and the transmission fluctuations  
\begin{equation}
\delta T_p {\,=\,} \sqrt{\overline{(T_p-\overline{T_p})^2}} \quad, \label{deltaTp}
\end{equation}
can be compared with $\xi_{loc}$~. The result is shown in Fig.\ref{Fig-SMP-AVE} for phase fluctuations following the u-G model [panel a)] and the c-LC model [panel b)]. As one can see, $\overline{T_p}$ (black solid line) and $T_p^{typ}$ (red dashed line) are different (in particular $T_p^{typ} <\overline{T_p}$), a typical signature that 1D disordered systems are not self-averaging.~\cite{kramer-review} Furthermore, the fluctuations $\delta T_p$ (blue dotted line) are of the order of the average transmission and, as expected for mesoscopic samples, do not decrease with an increasing number of samples. For the u-G model, all these three quantities follow the `Fraunhofer-like' energy pattern of $\xi_0/\xi_{loc}(E)$,  shown in Fig.\ref{Fig-xi-loc} and given in Fig.~\ref{Fig-SMP-AVE}a) as a thin black line as a guide to the eye. The transmission coefficient reaches 1 and its fluctuations $\delta T_p$ vanish at the energy values for which $\xi_{loc}$ diverges. For the c-LC model, the fluctuations $\delta T_p$ exhibit a local minimum at $E=0$, as shown in Fig.~\ref{Fig-SMP-AVE}b). Its origin can be understood in terms of the energy-shift effect $E\rightarrow E-\hbar v_F k_p^{(j)}$ that arises in this model and is discussed above. For a given value of ${\rm SD}(k_p)$, the energy ranges in which $T_p$ is less affected by such disorder-induced shift are  small energies $|E| \ll \Gamma_p^0$ and very high energies $|E| \gg \Gamma_p^0$, where the local transmission is likely to remain in the `sub-gap' region and in the `supra-gap' region, respectively, despite the fluctuating shift. This is why $\delta T_p$ are minimal in these ranges. When, however, the amount of the fluctuating energy shift becomes much bigger than the clean case `gap' $\Gamma_p^0$ (i.e. for $\hbar v_F {\rm SD}(k_p) \gg \Gamma_p^0$), the minimum at $E=0$ increases and the fluctuations $\delta T_p$ become energy independent.
We also notice that, while in a single sample  the phase fluctuations typically lead to a $T_p$ that is not particle-hole symmetric (see the red dashed curve in Fig.\ref{Fig-SGL-Gamma_1}), Fig.\ref{Fig-SMP-AVE} shows that particle-hole symmetry is recovered upon sample averaging.

%%%%%%%%%%%%%%%%%%%%%%%%%%%%%%%%%%
%%%%%%%%%%%%%%%%%%%%%%%%%%%%%%%%%%

%%%%%%%%%%%%%%%%%%%%%%%%%%%%%%%%%%%%%%%%%%%%%%%%%%%%%%%%% 
%%%%%%%%%%%%%%%%%%%%%%%%%%%%%%%%%%%%%%%%%%%%%%%%%%%%%%%%% 
%%%%%%%%%%      D I S C U S S I O N       %%%%%%%%%%%%%%%
%%%%%%%%%%%%%%%%%%%%%%%%%%%%%%%%%%%%%%%%%%%%%%%%%%%%%%%%% 
%%%%%%%%%%%%%%%%%%%%%%%%%%%%%%%%%%%%%%%%%%%%%%%%%%%%%%%%% 
\section{Discussion}
\label{sec-V}
\subsection{Geometrical {\it vs} electrical disorder}
We now want to discuss the previous results in terms of realistic systems, identifying situations where the three models of phase fluctuations can be applied. 
As observed in the Introduction, in realistic implementations of helical edge state  tunnel junctions, disorder has mainly two origins. One is the presence of roughness in the borders delimiting the constriction, which causes the width $W(x)$ of the junction to fluctuate along the longitudinal direction $x$. We shall refer to that as the `geometrical disorder'. The second origin is the presence of locally fluctuating potentials, due to  the oxides arising at the etching process and/or to the amorphous dielectric below the top gate. We shall refer to the latter as the `electrical disorder'.  

\begin{figure} 
\centering
\includegraphics[width=\columnwidth,clip]{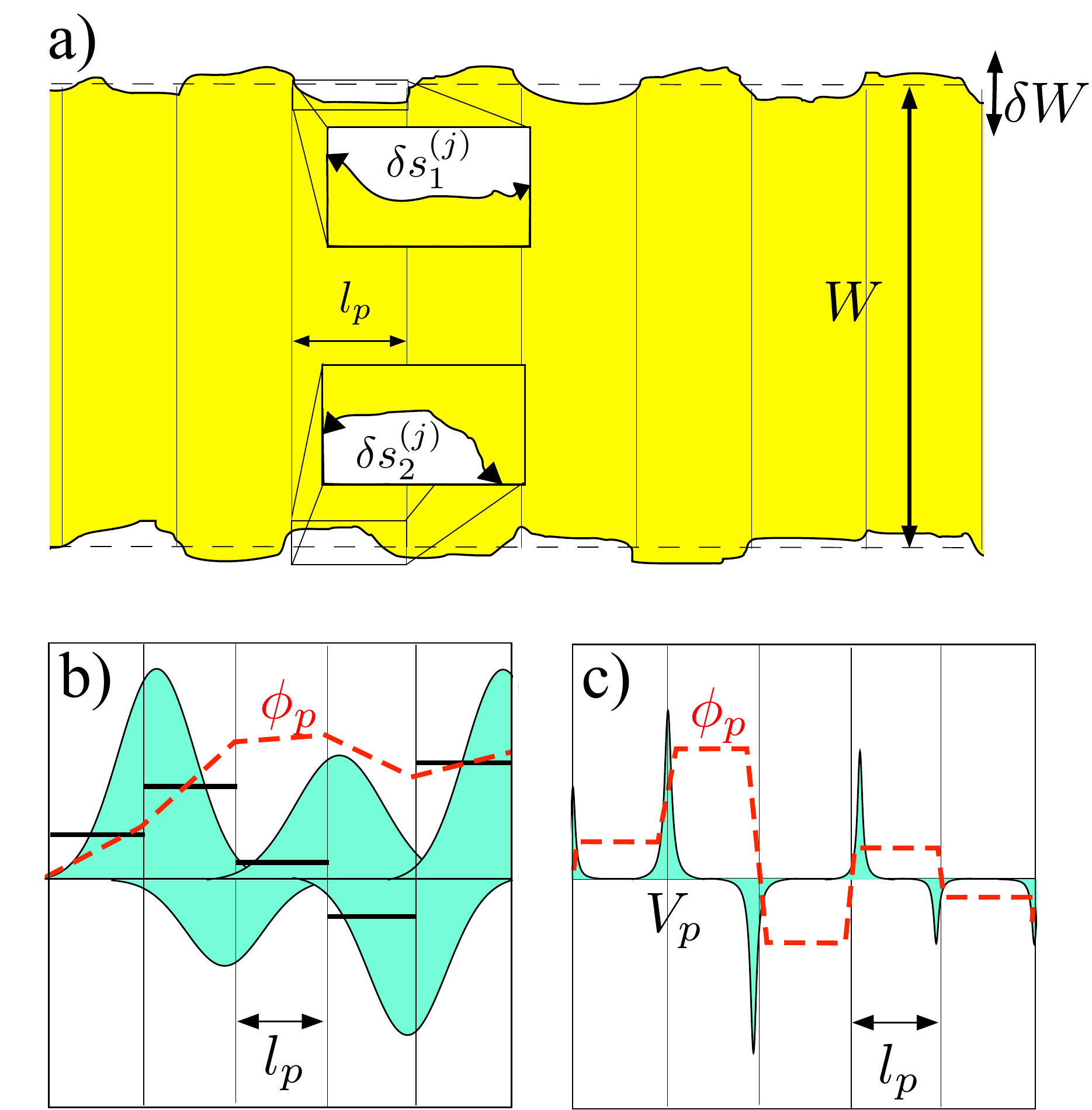} 
\caption{(Color online) Various sources of disorder in a helical tunnel junction: a) geometrical disorder and b) and c) electrical disorder. See text for details. For the geometrical disorder the fluctuations $\delta W$ of the width are typically smaller than the average width ($\delta W \ll W)$, so that $|\Gamma_p| \simeq \langle|\Gamma_p|\rangle$ is weakly disordered. However, the difference $\delta s^{(j)}=\delta s_1^{(j)}-\delta s_2^{(j)}$ between opposite profile lengths may cause fluctuations on the phase $\phi_p$ of the tunneling amplitude. In panels b) and c) the shaded area represents the typical profile of electrical disorder along the longitudinal direction of the helical tunnel junction. Panels b) and c) describe  the cases of long and short range potential, respectively, as compared to the typical potential center distance $l_p$. The red dashed curves represent  the equivalent formulation in terms of the phase of the tunneling amplitude, after applying the gauge transformation equation~(\ref{gauge-phase}).}
\label{Fig-geom-vs-elec}
\end{figure}

Concerning the geometrical disorder, it is worth mentioning that current lithographic techniques allow us to obtain an extremely precise profile. Thus, although some roughness is ultimately unavoidable due to lithographic resolution ($\sim 20 {\rm nm}$),  one can quite reasonably treat the local width $W(x)$ as a fluctuating variable that is sharply peaked around its average value $W \sim 100-300 \, {\rm nm}$ [see Fig.\ref{Fig-geom-vs-elec}a)]. Because $W(x)$ determines the local absolute value $|\Gamma_p(x)|$ of the tunneling amplitude, one can fairly claim that the distribution of $|\Gamma_p(x)|$ is also sharply peaked around its average value. The Rayleigh distribution (i.e. the distribution of the absolute value of a complex Gaussian variable) is not particularly peaked, for its  standard deviation is proportional to its average value,   ${\rm SD}(|\Gamma_p|) \simeq 0.52 \, \langle |\Gamma_p|\rangle$, and it overestimates the actual amount of fluctuations of $|\Gamma_p(x)|$. 
Nevertheless, even when the fluctuations of $|\Gamma_p(x)|$ are assumed to follow such a distribution, the transmission coefficient turns out to be essentially unaltered with respect to the clean case, as we have shown in Sec.\ref{sec-III}. We conclude that, for practical purposes, one can fairly approximate the absolute value as being locked to its average value, $|\Gamma_p(x)|\equiv \Gamma_p^0$, corresponding to the sharply peaked average width. 
To estimate the amount of phase fluctuations, one observes that the local tunneling amplitude  $\Gamma_p(x)$ is related to the overlap integral between the two uncoupled edge states wavefunctions $\Phi(x,y)$ over a local randomly varying area $A(x) \sim l_p \, W(x)$ centered around a longitudinal point $x$ over the typical roughness lengthscale $l_p$. Because the wavefunctions are of the form $\Phi(x,y)=e^{i k x} \varphi(y)$ where $\varphi(y)$ is a {\it real} function that transversally decays, the only contribution to   phase fluctuations is roughly $\Delta \phi_p^{(j)} \sim k \, \delta s^{(j)}$, where $\delta s^{(j)}$ is the difference between the lengths of the two edge profiles around~$x$ [see Fig.\ref{Fig-geom-vs-elec}a)]. For the helical states the Dirac spectrum   is centered around $k=0$, so that for a 1 m{\rm eV} energetic electron one has $k \sim 10^6 {\rm m}^{-1}$, while the roughness $\delta s$ is limited by a lithographic resolution of 20 nm. It is thus reasonable to assume that $\Delta \phi_p^{(j)} < 1$ and that the  phase fluctuations due to geometrical disorder are fairly small. They can be accounted for by either the c-RW model (with a parameter $\Delta \phi_p <1$) or  the c-LC model [with a parameter ${\rm SD}(k_p) \, l_p <1$]. In this limit small step-like variations and linear variations with a small slope are equivalent.  For the above reasons, we believe that in realistic junctions geometrical disorder due to roughness does not play a major role.

For the electrical disorder the scenario is richer. Let us consider the random potential $V_{p}(x)$, which consists of the sum of various electric potential sources, and denote by~$\lambda_{p}$ the typical extension range of these potential centers, and by $l_p$ their average distance from each other (notice that the value of $l_p$ for the electrical disorder can of course differ from the roughness lengthscale mentioned above for the geometrical disorder).
There can be two limiting configurations, depicted in panels b) and c) of Fig.\ref{Fig-geom-vs-elec}: in the case with $\lambda_{p}\gtrsim l_p$ [panel b)], the average of the potential over the distance $l_p$ is fairly representative of the potential distribution at such a lengthscale. Thus, in defining a coarse grained disorder one can assign a constant value of the potential to the interval (denoted by black thick lines). In contrast, in the opposite regime $\lambda_{p} \ll l_p$ [panel c)] the inhomogeneous distribution of strongly peaked potential centers cannot be replaced by the averaged potential over $l_p$. However, this difference can be reworded in terms of the phase $\phi_p$ of the tunneling amplitude. Indeed, as shown in Sec.\ref{Sec-IV-a},  with the gauge transformation  (\ref{trans-prime}) one can eliminate the potential $V_p(x)$ by introducing a renormalized phase $\phi_p \rightarrow \phi^\prime_p$ of the tunneling amplitude, given by the integral of $V_{p}$ [see Eq.(\ref{gauge-phase})]. Thus,   in the first case $\lambda_{p}\gtrsim l_p$ one obtains a linearly  varying phase [dashed red line in panel b)], and the c-LC model applies. 
In the second case, $\lambda_{p} \ll l_p$, the phase exhibits jumps at the potentials source centers [dashed red line in panel c)]. The amount of such phase jumps depends on the typical strength  of the potential peaks, $\Delta \phi_p \sim e\int_{l_p}dx V_p(x)/\hbar v_F$. The situation of sharp and strong peaks ($\Delta \phi_p \sim 2\pi$) can be described by the u-G model, whereas for weaker peaks one can simply replace Eq.(\ref{phase-u-G}) for the u-G model by a uniform distribution with $\Delta \phi_p < 2\pi$, similarly to what has been done in the analysis of Fig.\ref{Fig-SGL-Gamma_2}.\\

%%%%%%%%%%%%%%%%%%%%%%%%%%%%%%%%%%%%%%%%%%%%%%%%%%%
%%%%%%%%%%%%%%%%%%%%%%%%%%%%%%%%%%%%%%%%%%%%%%%%%%%

{\it Experimental conditions}.  Tunnel junctions in QSHE can be realized by lateral etching of HgTe/CdTe~\cite{TM-1-exp} and InAs/GaSb~\cite{InAsGaSb} quantum wells, and lithographic techniques can be exploited to tailor arbitrary shapes. The presence of the top gate enables one to enter the QSHE regime and to tune the Fermi level $E_F$. For constriction widths $W \sim 100-200 \, {\rm nm}$ in the HgTe/CdTe case, one obtains a tunneling amplitude magnitude $|\Gamma_p^0| \sim 0.25 - 2.5 \, {\rm meV}$.~\cite{zhou,richter,citro-sassetti}  These energy values are well below the bulk gap and are experimentally accessed.~\cite{TM-1-exp} Using the value of the Fermi velocity $v_F \simeq 0.5 \times   10^6 \, {\rm m/s}$,~\cite{zhang-review} a  lengthscale  $\xi_0=\hbar v_F/|\Gamma_p^0| \sim 0.1 - 1 \,\mu{\rm m}$ is obtained. For InAs/GaSb quantum wells, smaller values for $\xi_0$ may be expected, due to a Fermi velocity $v_F \simeq 2 \times   10^4 \, {\rm m/s}$ that is more than one order of magnitude smaller than that in HgTe/CdTe edge states.~\cite{InAsGaSb}
Thus,  for an $L \sim  1 \mu{\rm m}$ long junction, one has $\xi_0/L \sim 0.1-1$. The geometrical disorder and the electrical disorder discussed above thus occur over a typical lengthscale $l_p < \xi_0$, whose value depends on the specific fabrication method.  \\

%%%%%%%%%%%%%%%%%%%%%%%%%%%%%%%%%%%%%%%%%%%%%%%%%%%%%%%%%%%%%
%%%%%%%%%%%%%%%%%%%%%%%%%%%%%%%%%%%%%%%%%%%%%%%%%%%%%%%%%%%%%
%%%%%%%%%%%%%%%%%%%%%%%%%%%%%%%%%%%%%%%%%%%%%%%%%%%%%%%%%%%%%
%%%%%%%%%%%%%%%%%%%%%%%%%%%%%%%%%%%%%%%%%%%%%%%%%%%%%%%%%%%%%
%%%%%%%%%%%%%%%%%%%%%%%%%%%%%%%%%%%%%%%%%%%%%%%%%%%%%%%%%%%%%

\subsection{Differences from a disordered quantum wire}
Formal analogies arise between the disordered tunnel junction of helical states and the problem of a 1D disordered quantum wire. Indeed the $\Gamma_p(x)$ term and the $V_p(x)$ term play the role of backward scattering (BS) and forward scattering (FS) in a disordered wire, respectively.  

There are, however, various aspects that distinguish the problem of a tunnel junction of helical edge states from the case of a quantum wire. 
In the first instance, far away from the constriction region the helical states eventually separate, enabling one to separately measure the backscattering and the transmitted currents, given by Eqs.(\ref{I1}) and (\ref{I3}), respectively, for an injection from e.g. terminal 2.\cite{nota-T4}  This separation is not possible in a quantum wire.

Secondly, while in a quantum wire impurities are typically distributed everywhere, here disorder is effective over only the finite length $L$ of the constriction, because topological protection from disorder occurs away from the tunnel junction.

Furthermore, in quantum wires the disorder effects should be compared with the disorder-free case, which is a clean conducting wire with a roughly energy independent transmission coefficient. 
As a consequence,  only one typical electron lengthscale can compared with the disorder fluctuation lengthscale, namely, the Fermi wavelength $\lambda_F$ of a traveling wave in the clean case at the Fermi energy, which is of the order a few angstroms. 
In contrast, a clean tunnel junction with a uniform tunneling $\Gamma_p(x) \equiv \Gamma_p^0 \exp[i \phi_p]$  exhibits a much richer structure. Indeed one can identify the `sub-gap' region $|E|<\Gamma_p^0$, characterised by decaying electron waves  with a wavelength that takes the minimal value $\xi_0=\hbar v_F/\Gamma_p^0$ at $E=0$ and diverges at $E=\Gamma_p^0$, and the `supra-gap' region $|E|>\Gamma_p^0$, characterised by traveling waves with a wavelength decreasing in energy down to a behavior $\lambda_E \simeq \hbar v_F/E$ for high energies. This causes the energy dependence of the clean-case localization length Eq.(\ref{xi-cl}). 
When investigating the disorder effects in a tunnel junction, such rich energy dependence  of the clean case becomes crucial.  

Finally,  in a disordered quantum wire, FS and BS terms have the  {\it same} physical origin, namely the presence of impurities in the wire, and are associated with the $k \simeq 0$ and $k\simeq 2k_F$ Fourier components of the impurity potential, respectively, so that disorder always involves {\it both}  terms. In contrast, in a tunnel junction $\Gamma_p(x)$ is mainly due to the geometrical disorder and $V_p(x)$ is due to the electrical disorder, and we have argued above that the former disorder is significantly less relevant than the latter. 
This difference has severe physical implications, which can be described by invoking the gauge transformation (\ref{trans-prime}) that casts the FS term  $V_p$ into the renormalized phase $\phi^\prime_p$ [see Eq.(\ref{gauge-phase})] of a new BS term $\Gamma^\prime_p=|\Gamma_p| \exp[i \phi^\prime_p]$.  
Now, {\it if} the original $\Gamma_p$ is a Gaussian-distributed BS term, then the new $\Gamma^\prime_p$ obtained after the gauge  is {\it also} a Gaussian distributed variable. The FS term  is thus effectively canceled by the local correlation of the original BS term $\overline{\Gamma_p(x) \Gamma_p^*(x^\prime)} = g_0 \delta(x-x^\prime)$, and does not play any role. This is indeed the case of a quantum wire.~\cite{giamarchi}
However, if the original BS can be considered to be {\it not} disordered, the FS term {\it indirectly} induces a fluctuating BS term  $\Gamma^\prime_p$,  via Eq.(\ref{trans-prime}). The non-locality of the gauge transformation (\ref{gauge-phase}) maps an {\it uncorrelated} disorder on the FS term $V_p$, $\overline{V_p(x) V_p^*(x^\prime)} = v_0 \delta(x-x^\prime)$, into a {\it correlated non-Gaussian} disorder on the BS term $\Gamma_p$, whose space-correlations are related to the distribution parameter $v_0$ of the FS term~$V_p$.  This effect causes the strikingly different behaviors of the localization length in the u-G and the c-LC models, described by the black solid curve and the red dashed curve in Fig.\ref{Fig-xi-loc}. Notice that this phenomenon is essentially different from the more customary case of the Anderson model with correlated disorder, where correlations are introduced directly on the disordered potential~\cite{demoura,titov-schomerus,oliveira,croy,nguyen-kim}, and is more similar to the case of periodic-on-average systems.\cite{lisyansky}

%%%%%%%%%%%%%%%%%%%%%%%%%%%%%%%%%%%%%%%%%%%%%%%%%%%%%%%%%%
%%%%%%%%%%%%%%%%%%%%%%%%%%%%%%%%%%%%%%%%%%%%%%%%%%%%%%%%%%
%%%%%%%%%%%%%%%%%%%%%%%%%%%%%%%%%%%%%%%%%%%%%%%%%%%%%%%%%%
%%%%%%%%%%%%%%%%%%%%%%%%%%%%%%%%%%%%%%%%%%%%%%%%%%%%%%%%%%
%%%%%%%%%%%%%%%%%%%%%%%%%%%%%%%%%%%%%%%%%%%%%%%%%%%%%%%%%%
%%%%%%%%%%%%%%%%%%%%%%%%%%%%%%%%%%%%%%%%%%%%%%%%%%%%%%%%%%
\section{Conclusions}
We have analyzed the effects of disorder on a tunnel junction of helical edge states (Fig.\ref{Fig-01}) using an effective 1D model where the tunneling amplitude $\Gamma_p(x)=|\Gamma_p(x)| \exp[i \phi_p(x)]$ is randomly varying along the longitudinal direction of the junction. The analysis of the transmission coefficient $T_p$ of individual samples has shown that the disorders of the absolute value $|\Gamma_p|$ and of the phase $\phi_p$ of the tunneling amplitude lead to quite different effects,  the latter being typically more relevant in the physical regime $l_p < \xi_0 < L$, as shown in Fig.\ref{Fig-SGL-Gamma_1}. In particular, in the regime $l_p \ll \xi_0 < L$,  the fluctuations of  $\phi_p$ can even suppress tunneling  strongly, leading to an energy independent transmission coefficient (see Fig.\ref{Fig-SGL-Gamma_2}). This is in striking contrast to the behavior of a clean junction, where the transmission coefficient is independent of the value of the phase.\\
Furthermore, we have also shown that, when phase fluctuates by an amount of about $\Delta\phi_p \sim 2\pi$,  the way the phase fluctuates also becomes important. In particular, we have analyzed three different models for phase fluctuations (see Fig.\ref{Fig-phase-models}), and we have shown that the energy dependence of the localization length $\xi_{loc}(E)$ is dramatically different for the three cases, as illustrated in Fig.\ref{Fig-xi-loc}. Furthermore, the sample-averaged transmission, displayed in Fig.\ref{Fig-SMP-AVE},  turns out to be different.
Finally, we have discussed the physical situations where these three models apply in realistic tunnel junction implementations, and we have outlined the differences with respect to disordered quantum wires.

\acknowledgments
The authors greatly acknowledge D. Basko, G. Tkachov, B. Trauzettel, F. Cr\'epin, and H. Buhmann for fruitful and inspiring discussions. F.D. also  acknowledges financial support from Gastprofessorenprogramm 2013 Universit\"at W\"urzburg, and from Italian FIRB 2012 project HybridNanoDev (Grant No.RBFR1236VV).

%%%%%%%%%%%%%%%%%%%%%%%%%%%%%%%%%%%%%%%%%%%%%%%%%%%%%%%
%%%%%%%%%%%%%%%%%%%%%%%%%%%%%%%%%%%%%%%%%%%%%%%%%%%%%%%
%%%%%%%%%%%%%%%%%%%%%%%%%%%%%%%%%%%%%%%%%%%%%%%%%%%%%%%
%%%%%%%%%%%%%%%%%%%%%%%%%%%%%%%%%%%%%%%%%%%%%%%%%%%%%%%
%%%%%%%%%%%%%%%%%%%%%%%%%%%%%%%%%%%%%%%%%%%%%%%%%%%%%%%
%%%%%%%%%%%%%%%%%%%%%%%%%%%%%%%%%%%%%%%%%%%%%%%%%%%%%%%
%%%%%%%%%%%%%%%%%%%%%%%%%%%%%%%%%%%%%%%%%%%%%%%%%%%%%%%
%%%%%%%%%%%%%%%%%%%%%%%%%%%%%%%%%%%%%%%%%%%%%%%%%%%%%%%
%%%%%%%%%%%%%%%%%%%%%%%%%%%%%%%%%%%%%%%%%%%%%%%%%%%%%%%
\appendix
\section{Transfer Matrix for the three phase disorder models}
\label{App-A}
\noindent The tunnel region is divided into $N$ intervals $[x_{j-1};x_{j}]$ ($j=1,\ldots N$), characterised by the disorder lengthscale $l_p$ each. Here $x_0$ and $x_N=x_f$ are the left and right extremal points of the tunnel region, respectively. Within each interval the absolute value of the tunneling amplitude is assumed to take a constant value 
\begin{equation}
|\Gamma_p(x) |\equiv |\Gamma_p^{(j)}| \hspace{1cm} x_{j-1} \le  x < x_j \quad.
\end{equation} 
The phase $\phi_p(x)$  fluctuates from one interval to another according to the three different models described in Sec.\ref{Sec-IV-a}.  
The transfer matrix of the sample is obtained from the evolution operator through Eq.(\ref{mp-def}), where the evolution operator $\mathbf{U}_{p,E}(x_f;x_0)$ is provided here below for the three phase fluctuation models:\\

\noindent i) models u-G and c-RW\\
For the u-G and c-RW models the phase takes a constant value $\phi_p^{(j)}$ within each interval, given by Eqs.(\ref{phase-u-G}) and (\ref{phase-c-RW}), respectively. The total evolution operator across the junction is given by
\begin{equation}
\mathbf{U}^{\mu}_{p,E}(x_f;x_0) = \prod_{j=N}^1 \mathbf{U}^{\mu}_{p,E}(x_j;x_{j-1}) \hspace{0.5cm}  \label{U-c-uG_1}  \quad,  
\end{equation}
where $x_N=x_f$ and  $\mu=\mbox{u-G, c-RW}$, and
\begin{eqnarray}
\lefteqn{\mathbf{U}^{\mu}_{p,E}(x_{j};x_{j-1}) =} \hspace{1cm} & &  \nonumber \\ 
& & =  \,  {\rm T} \, e^{-i \int_{x_{j-1}}^{x_j} \, \boldsymbol\tau \cdot  \mathbf{b}_{p,E}^{(j)}(x^\prime)  \, dx^\prime}   =   e^{-i  \, l_p \boldsymbol\tau \cdot  \mathbf{b}_{p,E}^{(j)}   \,   }    \nonumber   \\
 & &   \,    \nonumber    \\
& & =  \left\{ \begin{array}{l}
 \tau_0 \cos(\tilde{k}_E^{(j)} l_p)\,  -i  \boldsymbol\tau \cdot \mathbf{b}_{p,E}^{(j)}   \frac{\sin(\tilde{k}_E^{(j)}  l_p) }{\tilde{k}_E^{(j)}}      \,  \\ \\  \hspace{3.5cm} \mbox{for} \, \,  |E|>|\Gamma_{p}^{(j)}| \\ \\ \\
 \tau_0\cosh(\tilde{q}_E^{(j)}  l_p) \,  - i \boldsymbol\tau \cdot \mathbf{b}_{p,E}^{(j)}  \frac{\sinh(\tilde{q}^{(j)}_E l_p) }{ \tilde{q}_E^{(j)} }   \, \\ \\
   \hspace{3.5 cm} \mbox{for} \, \, |E|<|\Gamma_{p}^{(j)}| \end{array}\right.
 \label{U-c-uG_2}
\end{eqnarray}
is and the evolution operator in the $j$-th interval. Here
\begin{eqnarray}
\label{tildekE_models-i}
\begin{array}{lcl} \tilde{k}^{(j)}_E = \frac{\sqrt{ E^2-|\Gamma_{p}^{(j)}|^2}}{\hbar v_F} & \mbox{for} & |E|>|\Gamma_{p}^{(j)}| \\ & & \\ \tilde{q}_E^{(j)} =\frac{\sqrt{|\Gamma_{p}^{(j)}|^2-E^2}}{\hbar v_F}  &  \mbox{for} & |E|<|\Gamma_{p}^{(j)}|\end{array}  
\end{eqnarray}
and 
\begin{eqnarray}
 \mathbf{b}^{(j)}_{p,E}   =  \frac{(-i\,  |\Gamma_{p}^{(j)}|  \sin  \phi_{p}^{(j)}   , \,  i\, |\Gamma_{p}^{(j)}|  \cos  \phi_{p}^{(j)}(x)     \, , \,   -E)}{\hbar v_F} \quad.\nonumber 
\end{eqnarray}

\noindent ii) model c-LC\\
For this model the phase $\phi_p(x)$ varies  linearly within each interval  according to Eq.(\ref{phase-c-LC}). The related $\mathbf{b}^{\rm c-LC}_{p,E}(x)$ vectors in Eq.(\ref{bpE-def})  oscillate in space and, as a consequence, the matrices $\boldsymbol\tau \cdot \mathbf{b}^{\rm c-LC}_{p,E}(x)$ and $\boldsymbol\tau \cdot \mathbf{b}^{\rm c-LC}_{p,E}(x^\prime)$ at any two different points {\it do not commute}, even within the same interval~$j$, so that
\begin{equation}
{\rm T} \, e^{-i \int_{x_{j-1}}^{x_j} \! \boldsymbol\tau \cdot  \mathbf{b}_{p,E}^{{\rm c-LC}(j)}(x^\prime)  \, dx^\prime}   \neq    e^{-i \int_{x_{j-1}}^{x_j} \! \boldsymbol\tau \cdot  \mathbf{b}_{p,E}^{{\rm c-LC}(j)}(x^\prime)  \, dx^\prime} \quad.
\end{equation}
This makes the evaluation of the time-ordered exponential in (\ref{Up-expr}) {\it a priori} non-trivial. To circumvent this problem, we observe that the linearly continuous behavior (\ref{phase-c-LC}) can be rewritten as
\begin{equation}
\phi^{\rm c-LC}_p(x)=\phi_p^0 -2 \int_{x_0}^x k_{p}(x^\prime) \, dx^\prime \label{phase-cLc-bis}
\end{equation}
where $\phi_p^0 $ is a constant and $k_{p}(x)$ is a random piece-wise constant function
\begin{equation}
k_{p}(x) \equiv  k_p^{(j)}  \hspace{1cm} x_{j-1} \le  x < x_j \quad.
\end{equation} 
The phase (\ref{phase-cLc-bis}) can thus be regarded to as the   result Eq.(\ref{gauge-phase}) of  an applied gauge transformation (\ref{gauge-trans}), upon setting $\phi^\prime_p(x)=\phi^{\rm c-LC}_p(x)$  and $V_{p}(x)=\hbar v_F k_{p}(x)/e$. Then, the evolution operator for the c-LC model is easily written [see Eq.(\ref{Uptilde-expr})] as  
\begin{eqnarray}
\lefteqn{\mathbf{U}^{\rm c-LC}_{p,E}(x_f;x_0)= } \hspace{1cm} & & \nonumber \\
& & =e^{i \int_{x_0}^{x_N} k_{p}(x^\prime) \tau_z}  \prod_{j=N}^1 \mathbf{U} _{p,E}(x_j;x_{j-1})  \label{U-c-LC_1}   \quad,  
\end{eqnarray}
with
\begin{eqnarray}
\lefteqn{\mathbf{U}_{p,E}(x_{j};x_{j-1})=     } \hspace{1cm} & &  \nonumber \\
 & &   \,    \nonumber    \\
& &=   \left\{ \begin{array}{l}
 \tau_0 \cos(\tilde{k}_E^{(j)} l_p)\,  -i  \boldsymbol\tau \cdot \mathbf{b}_{p,E}^{(j)}   \frac{\sin(\tilde{k}_E^{(j)}  l_p) }{\tilde{k}_E^{(j)}}      \,  \\ \\  \hspace{2.5cm} \mbox{for} \, \,  |E-\hbar v_F k_{p}^{(j)}|>|\Gamma_{p}^{(j)}| \\ \\ \\
 \tau_0\cosh(\tilde{q}_E^{(j)}  l_p) \,  - i \boldsymbol\tau \cdot \mathbf{b}_{p,E}^{(j)}  \frac{\sinh(\tilde{q}^{(j)}_E l_p) }{ \tilde{q}_E^{(j)} }    \, \\ \\
   \hspace{2.5 cm} \mbox{for} \, \, |E-\hbar v_F k_{p}^{(j)}|<|\Gamma_{p}^{(j)}| \end{array}\right.
 \label{U-c-LC_2} 
\end{eqnarray}
Here 
\begin{eqnarray}
\label{tildekE_models-ii}
\begin{array}{lcl} \tilde{k}^{(j)}_E = \frac{\sqrt{(E-\hbar v_F k_{p}^{(j)})^2-|\Gamma_{p}^{(j)}|^2}}{\hbar v_F} & \mbox{for} & |E-\hbar v_F k_{p}^{(j)}|>|\Gamma_{p}^{(j)}| \\ & & \\ \tilde{q}_E^{(j)} =\frac{\sqrt{|\Gamma_{p}^{(j)}|^2-(E-\hbar v_F k_{p}^{(j)})^2}}{\hbar v_F}  &  \mbox{for} & |E-\hbar v_F k_{p}^{(j)}|<|\Gamma_{p}^{(j)}|\end{array}  
\end{eqnarray}
are the local wavevectors and 
\begin{eqnarray}
 \mathbf{b}_{p,E}^{(j)}   =\frac{(-i\,  |\Gamma_{p}^{(j)}|  \sin  \phi_{p}^{0}     , \,  i\, |\Gamma_{p}^{(j)}|  \cos  \phi_{p}^{0}      \, , \,  \hbar v_F k_p^{(j)} -E)}{\hbar v_F}\nonumber  
\end{eqnarray}
are the local `magnetic' fields that the gauge transformation would transform into the $\mathbf{b}_{p,E}^{{\rm c-LC}(j)}$ fields.

%%%%%%%%%%%%%%%%%%%%%%%%%%%%%%%%%%%%%%%%%%%%%%%%%%%%%%%%%%%%%%%%%%%%%%%%%%%%%%%%%%%%
%%%%%%%%%%%%%%%%%%%%%%%%%%%%%%%%%%%%%%%%%%%%%%%%%%%%%%%%%%%%%%%%%%%%%%%%%%%%%%%%%%%%
%%%%%%%%%%%%%%%%%%%%%%%%%%%%%%%%%%%%%%%%%%%%%%%%%%%%%%%%%%%%%%%%%%%%%%%%%%%%%%%%%%%%
%%%%%%%%%%%%%%%%%%%%%%%%%%%%%%%%%%%%%%%%%%%%%%%%%%%%%%%%%%%%%%%%%%%%%%%%%%%%%%%%%%%%
%%%%%%%%%%%%%%%%%%%%%%%%%%%%%%%%%%%%%%%%%%%%%%%%%%%%%%%%%%%%%%%%%%%%%%%%%%%%%%%%%%%%
%%%%%%%%%%%%%%%%%%%%%%%%%%%%%%%%%%%%%%%%%%%%%%%%%%%%%%%%%%%%%%%%%%%%%%%%%%%%%%%%%%%%
%%%%%%%%%%%%%%%%%%%%%%%%%%%%%%%%%%%%%%%%%%%%%%%%%%%%%%%%%%%%%%%%%%%%%%%%%%%%%%%%%%%%
%%%%%%%%%%%%%%%%%%%%%%%%%%%%%%%%%%%%%%%%%%%%%%%%%%%%%%%%%%%%%%%%%%%%%%%%%%%%%%%%%%%%
%%%%%%%%%%%%%%%%%%%%%%%%%%%%%%%%%%%%%%%%%%%%%%%%%%%%%%%%%%%%%%%%%%%%%%%%%%%%%%%%%%%%
%%%%%%%%%%%%%%%%%%%%%%%%%%%%%%%%%%%%%%%%%%%%%%%%%%%%%%%%%%%%%%%%%%%%%%%%%%%%%%%%%%%%
%%%%%%%%%%%%%%%%%%%%%%%%%%%%%%%%%%%%%%%%%%%%%%%%%%%%%%%%%%%%%%%%%%%%%%%%%%%%%%%%%%%%
%%%%%%%%%%%%%%%%%%%%%%%%%%%%%%%%%%%%%%%%%%%%%%%%%%%%%%%%%%%%%%%%%%%%%%%%%%%%%%%%%%%%
%%%%%%%%%%%%%%%%%%%%%%%%%%%%%%%%%%%%%%%%%%%%%%%%%%%%%%%%%%%%%%%%%%%%%%%%%%%%%%%%%%%%
%%%%%%%%%%%%%%%%%%%%%%%%%%%%%%%%%%%%%%%%%%%%%%%%%%%%%%%%%%%%%%%%%%%%%%%%%%%%%%%%%%%%
%%%%%%%%%%%%%%%%%%%%%%%%%%%%%%%%%%%%%%%%%%%%%%%%%%%%%%%%%%%%%%%%%%%%%%%%%%%%%%%%%%%%
%%%%%%%%%%%%%%%%%%%%%%%%%%%%%%%%%%%%%%%%%%%%%%%%%%%%%%%%%%%%%%%%%%%%%%%%%%%%%%%%%%%%
%%%%%%%%%%%%%%%%%%%%%%%%%%%%%%%%%%%%%%%%%%%%%%%%%%%%%%%%%%%%%%%%%%%%%%%%%%%%%%%%%%%%


\begin{thebibliography}{99}
\bibitem{footnote-RL-pm}  The notation $\alpha=R/L=\pm$ indicates that $\alpha=+$ for right(R) movers, and $\alpha=-$ for left(L) movers.
\bibitem{magnetic-fields}  The time-reversal (TR) transformation is $\mathcal{T}=(i\sigma_y \otimes \tau_x) K$, where $K$ denotes complex conjugation. Notice that, although the combinations $\boldsymbol\tau \cdot \mathbf{b}_{p,E}$ and $\boldsymbol\sigma \cdot \mathbf{b}_{f}$ are odd under TR, the way they appear in the Hamiltonian~(\ref{HAM-FER-MAT}) does preserve TR, due to the presence of the additional $\tau_z$ matrix acting in the chirality space.
\bibitem{nota-T4} For a comparison with the quantum wire case, one should set $\Gamma_f=0$, so that $T_{f}=1$, and no current flows into terminal~4.
\end{thebibliography}
\end{document}